%% file: submit.tex
\documentclass[preprint,review,3p,12pt]{elsarticle}

\usepackage{algorithm} \usepackage[noend]{algorithmic}
\usepackage{amsmath}
\usepackage{algorithm}
\usepackage[noend]{algorithmic}
\usepackage{bm}
\usepackage{amssymb}
\usepackage{stmaryrd}
\usepackage{multicol}
\usepackage{multirow}
\usepackage{epsfig}
\usepackage{subfigure}
\usepackage{graphicx}

\usepackage{amsthm}

\journal{Journal of Computational Physics}

\input{thesis_macro.tex}

\ifx\pdfoutput\undefined
\usepackage{graphicx}
\else
\usepackage{epstopdf}
\fi 

\begin{document}

\begin{frontmatter}

\title{Multibody Multipole Methods}

\author[label1]{Dongryeol Lee}

\address[label1]{Computational Science and Engineering\\Georgia Institute of
Technology\\266 Ferst Drive\\Atlanta, GA 30332}

\author[label2]{Arkadas Ozakin}

\address[label2]{Georgia Tech Research Institute\\Georgia Institute
of Technology\\266 Ferst Drive\\Atlanta, GA 30332}

\author[label1]{Alexander G. Gray}

\begin{abstract}
A three-body potential function can account for interactions among
triples of particles which are uncaptured by pairwise interaction
functions such as Coulombic or Lennard-Jones potentials. Likewise, a
multibody potential of order $n$ can account for interactions among
$n$-tuples of particles uncaptured by interaction functions of lower
orders.  To date, the computation of multibody potential functions for
a large number of particles has not been possible due to its $O(N^n)$
scaling cost. In this paper we describe a fast tree-code for
efficiently approximating multibody potentials that can be factorized
as products of functions of pairwise distances. For the first time, we
show how to derive a Barnes-Hut type algorithm for handling
interactions among more than two particles. Our algorithm uses two
approximation schemes: 1) a deterministic series expansion-based
method; 2) a Monte Carlo-based approximation based on the central
limit theorem. Our approach guarantees a user-specified bound on the
absolute or relative error in the computed potential with an
asymptotic probability guarantee.  We provide speedup results on a
three-body dispersion potential, the Axilrod-Teller potential.
\end{abstract}


\end{frontmatter}

\section{Introduction}
In this paper, we generalize previous algorithmic frameworks for
rapidly computing pair-wise summations to include higher-order
summations. Suppose we are given a set of particles $X = \{ x_0,
\cdots, x_{N - 1} \}$ in $D$-dimensional space. 

\begin{figure}[t]
\centering
\includegraphics[width=0.6\textwidth]{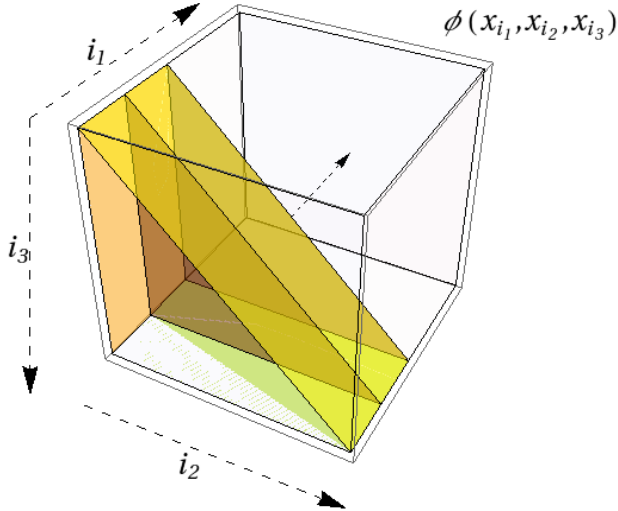}
\caption{An example multibody computation ($n = 3$). For each fixed
  argument $x_{i_1}$, $\Phi(x_{i_1})$ equals the summation of the
  entries $\phi(x_{i_1}, x_{i_2}, x_{i_3})$ in the shaded region
  corresponding to $x_{i_1}$.\label{fig:multibody:tensor_slice_sum}}
\end{figure}

For $x \in X$ and a $n$-tuple function $\phi: \underbrace{\mathbb{R}^D
  \times \cdots \times \mathbb{R}^D}_{n \mbox{ copies}} \rightarrow
\mathbb{R}$, we are interested in computing the following
form\footnote{In computing $\Phi(x)$, we fix one of the arguments of
  $\phi$ as $x$ and choose a $( n - 1 )$-subset from $X^{(n - 1)}$
  which does not contain $x$.}:
\begin{equation}
\Phi(x; \underbrace{X \times \cdots \times X}_{(n - 1)\mbox{ copies}}
) = \sum\limits_{ x_{i_2} \in X \backslash \{ x \} } \ \sum\limits_{
  \substack{x_{i_3} \in X \backslash \{ x \}\\ i_2 < i_3} } \cdots \sum\limits_{
 \substack{ x_{i_n} \in X \backslash \{ x \}\\ i_{n_1} < i_n} } \phi ( x, x_{i_2}
\cdots, x_{i_n} )
\label{eq:canonical_multibody_summation}
\end{equation}
Sums of the form Equation~\eqref{eq:canonical_multibody_summation}
occur in molecular dynamics, protein structure prediction, and other
similar contexts. Biomolecular simulations usually break down the
interactions in complex chemical systems into balls-and-springs
mechanical models augmented by torsional terms, pairwise point charge
electrostatic terms, and simple pairwise dispersion (van der Waals)
interactions, etc. However, such pairwise ($n = 2$) interactions often
fail to capture important, complex non-additive interactions found in
real systems.  Though many researchers have argued that multibody
potentials enable more accurate and realistic molecular modeling, the
evaluation of $n$-body forces for $n\ge 3$ in systems beyond tiny
sizes (less than 10,000 particles) has not been possible due to the
unavailability of an efficient way to realize the computation.

In this paper we focus on computing multibody potentials of the third
order ($n = 3$), but frame our presentation so that the
methods can easily be generalized to handle higher-order potentials. 
For concreteness, we consider the Axilrod-Teller potential
(dispersion potential):
\begin{equation}
\phi(x_i, x_j, x_k) = 
\frac{1 + 3 \cos{\theta_i} \cos{\theta_j}
\cos{\theta_k}}{||x_i - x_j||^3||x_i - x_k||^3 || x_j - x_k||^3}
\end{equation}
where $\theta_i$, $\theta_j$, $\theta_k$ are the angles at the
vertices of the triangle $x_i x_j x_k$ and $|| \cdot ||$ is the
Euclidean distance metric. This
potential~\cite{axilrod1943interaction} describes induced dipole
interactions between triples of atoms, and is known to be important
for the accurate computation of the physical properties of certain
noble gases. 

{\noindent \bf This Paper. }For the first time, we introduce a fast
algorithm for efficiently computing multibody potentials for a large
number of particles. We restrict the class of multibody potentials to
those that can be factorized as products of functions of pairwise
Euclidean distances. That is, { \small
\begin{equation}
\phi(x_{i_1}, \cdots, x_{i_n}) = \prod\limits_{1 \leq p < q \leq n}
\phi_{p, q}(x_{i_p}, x_{i_q}) = 
\prod\limits_{1 \leq p < q \leq n}
\phi_{p, q} (  x_{i_p} -  x_{i_q}  )
\label{eq:multibody_potential_form}
\end{equation}
}Our algorithm achieves speedup by utilizing two approximation
methods: a deterministic and a probabilistic one. The deterministic
approximation is based on the analytic series-expansion-based approach
in~\cite{cheng1999fam,greengard2002nvf,greengard1991fgt} to handle
potential functions that describe $n$-body interactions with $n>2$.
The probabilistic approach uses a Monte Carlo-based approximation
based on the central limit theorem.  Our algorithm can compute
multibody potentials within user-specified bounds for relative or
absolute error with an asymptotic probability guarantee. 

However, we would like to point out the following limitations in our
algorithm. First of all, we do not present a full-fledged derivation
of all three translation operators (namely the far-to-far, the
far-to-local, and the local-to-local translation operators) for the
general multibody case. While we define the far-field expansion for a
restricted class of multibody potentials, defining the local expansion
for this same class is harder (see
Section~\ref{sec:local_expansion_three_body_potential}). We would also
like to point out that the hybrid deterministic/probabilistic
approximation heuristic works under some partial distributions but not
all. Indeed, there are configurations for which the speedup factor
over the naive brute-force method is minimal. The Monte-Carlo based
approximation relies on two theorems: 1) the central limit theorem
from which we determine the number of required samples; 2) the
Berry-Esseen theorem which characterizes the the rate at which the
sample average converges to the true average. Both theorems provide
only asymptotic guarantees.

Our work utilizes and extends a framework for efficient algorithms for
so-called {\em generalized $N$-Body Problems}~\cite{gray2001nbp},
which introduced {\em multi-tree methods}. The framework was
originally developed to accelerate common bottleneck statistical
computations based on distances; it utilizes multiple $kd$-trees and
other spatial data structures to reduce computation times both
asymptotically and practically by multiple orders of magnitude.  This
work extends the framework with higher-order hierarchical series
approximation techniques, demonstrating a fast multipole-type method
for higher-order interactions for the first time, effectively creating
a \emph{Multibody Multipole Method}.

Section~\ref{sec:generalized_nbody_framework} introduces the {\em
  generalized $N$-body framework} and describes a partial extension of
fast multipole-type methods to handle higher-order interactions; we
will discuss the technical difficulties for deriving all of the
necessary tools for the general multibody case. As a result, we
utilize only a simple but effective approximation using the
center-of-mass approximations. Section~\ref{sec:three_body_algorithms}
focuses on three-body interactions and introduces methods to do potential
computations under both deterministic and probabilistic error
criteria; the section also provides a description of the fast algorithm for the
three-body case. Section~\ref{sec:correctness_of_the_algorithm} proves
that our proposed algorithms can approximate potentials within
user-specified error bounds. Section~\ref{sec:experimental_results}
shows experimental scalability results for our proposed algorithms
against the naive algorithm under different error parameter settings.

{\noindent \bf Notations. }Throughout this paper, we use these common sets of notations:
\begin{itemize}
\item{{\bf (Normal Distribution). }This is denoted by
  $\mathcal{N}(\mu, \Sigma)$ where $\mu$ and $\Sigma$ are the mean and the covariance respectively.}
\item{{\bf (Vector Component). }For a given vector $v \in
  \mathbb{R}^k$, we access its $d$-th component by ${v}[d]$
  where $1 \leq d \leq k$ (i.e. $1$-based index).}
\item{{\bf (Multi-index Notation\index{Multi-index notation}). }Throughout this paper, we will be
  using the multi-index notation. A $D$-dimensional multi-index
  $\boldsymbol\alpha$ is a $D$-tuple of non-negative integers and will
  be denoted using a bold lowercase Greek alphabet. For any
  $D$-dimensional multi-indices $\boldsymbol\alpha$, $\boldsymbol\beta$
  and any ${x} \in \mathbb{R}^D$,
\begin{itemize}
\item[]{$| \boldsymbol\alpha | = \boldsymbol\alpha [1] + \boldsymbol\alpha [2] + \cdots + \boldsymbol\alpha [D]$}
\item[]{$\boldsymbol\alpha! = (\boldsymbol\alpha[1])! (\boldsymbol\alpha[2])! \cdots (\boldsymbol\alpha[D])!$}
\item[]{${x}^{\boldsymbol\alpha} = ({x}[1])^{\boldsymbol\alpha[1]} ({x}[2])^{\boldsymbol\alpha[2]} \cdots
({x}[D])^{\boldsymbol\alpha[D]}$}
\item[]{$D^{\boldsymbol\alpha} = \partial_1^{\boldsymbol\alpha[1]} \partial_2^{\boldsymbol\alpha[2]}
\cdots \partial_D^{\boldsymbol\alpha[D]}$}
\item[]{$\boldsymbol\alpha + \boldsymbol\beta = (\boldsymbol\alpha[1] + \boldsymbol\beta[1], \cdots, \boldsymbol\alpha[D] +
\boldsymbol\beta[D])$}
\item[]{$\boldsymbol\alpha - \boldsymbol\beta = (\boldsymbol\alpha[1] - \boldsymbol\beta[1], \cdots, \boldsymbol\alpha[D] -
\boldsymbol\beta[D])$ for $\boldsymbol\alpha \geq \boldsymbol\beta$.}
\end{itemize}
where $\partial_i$ is a $i$-th directional partial
derivative. Define $\boldsymbol\alpha > \boldsymbol\beta$ if $\boldsymbol\alpha[d] > \boldsymbol\beta[d]$, and
$\boldsymbol\alpha \geq p$ for $p \in \mathbb{Z^+} \cup \{0 \}$ if $\boldsymbol\alpha[d]
\geq p$ for $1 \leq d \leq D$ (and similarly for $\boldsymbol\alpha \leq p$).}
\item{{\bf (Size of a Point Set). }Given a set $S$, it size
  is denoted by $| S | $.}
\item{{\bf (Probability Guarantee). }We use the unbold Greek alphabet $\alpha$.}
\item{{\bf (A Tree Node). }A tree node represents a subset of a point
  set represented by the root node. Hence, we use
  the same notation as the previous.}
\item{{\bf (Representative Point of a Tree Node). }Usually a geometric
  center is used but any point inside the bounding primitive of a
  tree node is chosen as well. For the tree node ${P}$, this is denoted as ${c_P}$.}
\item{{\bf (Child Nodes of an Internal Tree Node). }Given a node
  ${N}$, denote its left and right child nodes by
  ${N}^L$ and ${N}^R$ respectively.}
\end{itemize}

\section{Related Work}
\subsection{Error Bounds}
Due to its expensive computational cost, many algorithms approximate
sums at the expense of reduced precision. The following error bounding
criteria are used in the literature:
\begin{defn}
{\bf $\tau$ absolute error bound: }For each $\Phi(x)$ for $x \in X$, it
computes $\widetilde{\Phi}(x )$ such that $\left | \widetilde{\Phi}( x  ) - \Phi(x  ) \right |
\leq \tau$.
\label{defn:bound_absolute_error}
\end{defn}
\begin{defn}
{\bf $\epsilon$ relative error bound: }For each $\Phi ( x )$ for $x
\in X$, compute $\widetilde{\Phi}(x )$ such that $\left |
\widetilde{\Phi}(x ) - \Phi(x ) \right | \leq \epsilon \left | \Phi(x) \right |$.
\label{defn:bound_relative_error}
\end{defn}

Bounding the relative error is much harder because the error bound
criterion is in terms of the initially unknown exact quantity. As a
result, many previous methods~\cite{greengard1991fgt,yang2003improved}
have focused on bounding the absolute error. The relative error bound
criterion is preferred to the absolute error bound criterion in
statistical applications in which high accuracy is desired. Our
framework can enforce the following error form:
\begin{defn}
{\bf $(1-\alpha)$ probabilistic $\epsilon$ relative/$\tau$ absolute
  error: }For each $\Phi(x)$ for $x \in X$, compute $\widetilde{\Phi}(x )$, such
that with at least probability $0 < 1 - \alpha \leq 1$, $\left |
\widetilde{\Phi}(x ) - \Phi (x ) \right | \leq \epsilon \left | \Phi ( x  ) \right |+
\tau$.
\label{defn:bound_probabilistic_relative_error}
\end{defn}
\subsection{Series Expansion}A series of papers first laid the
foundations for efficiently computing sums of pairwise potentials such
as Coulombic and Yukawa
potentials~\cite{cheng1999fam,greengard2002nvf,greengard1991fgt}. The
common approach in these papers is to derive analytical series
expansions of the given potential function in either Cartesian or
spherical coordinate systems. The series expansion is then truncated
after taking a fixed number of terms. The associated error bounds are
derived from summing the truncated terms in an appropriate infinite
geometric sum or bounding the remainder term using Taylor's theorem. A
recent line of work on efficient computation of pairwise function has
focused on developing numerical representations of the potential
matrix $[\phi(x_m, x_n)]_{m, n = 1}^N$, rather than relying on
analytical expansion of the potential
function.~\cite{martinsson2008accelerated} and~\cite{kapur1998iee} use
singular value decomposition and the QR decomposition to compute the
compressed forms of the potential function and the three translation
operators.~\cite{anderson1992ifm,ying2004kia} take the
``pseudo-particle'' approach by placing equivalent artificial charges
on the bounding surface of the actual particles by solving appropriate
integral equations. All of these works have been limited to pairwise
potential functions, and the approach does not naturally suggest a
generalization to $n$-body potentials with $n>2$. To our knowledge, no
research has been performed on the problem of evaluating multibody
potentials using a method more sophisticated than the $O(N^n)$
brute-force algorithm with an ad-hoc cut-off
distance.~\cite{marcelli2001tro,marcelli2009beyond}.

\section{Generalized $N$-body Framework}
\label{sec:generalized_nbody_framework}
\begin{figure}[t]
\centering
\scalebox{0.5}{\includegraphics{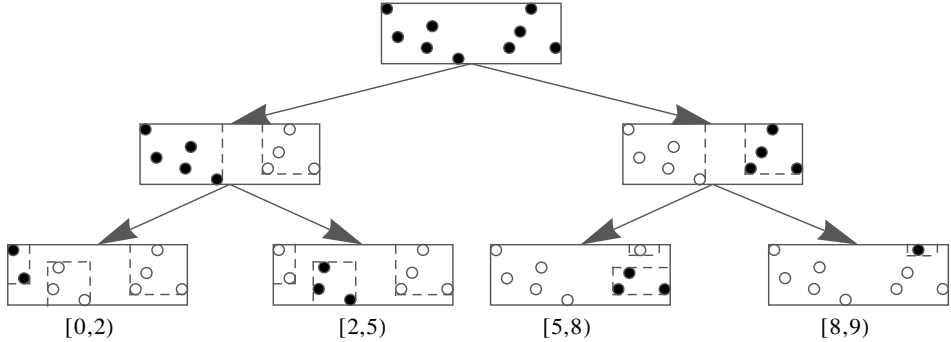}}
\caption{$kd$-tree of a two-dimensional point set. At each level, the
  bounding box is split in half along the widest dimension. The solid
  points denote the points owned by each node. At each leaf node, we
  can enumerate each point with its depth-first rank. The minimum
  depth-first rank (inclusive) and the maximum depth-first rank
  (exclusive) is shown for each
  node.\label{fig:multibody:kdtree_dfs_index}}
\end{figure}
\begin{figure}[t]
\subfigure[] {
\includegraphics[trim=0.25cm 1.1cm 0.25cm 0.25cm, clip=true]{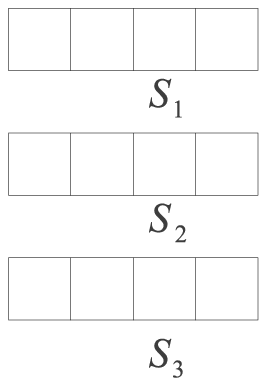}
\label{fig:num_tuples_canonical1}
}
\hspace{-3em}
\subfigure[] {
\includegraphics[trim=0.25cm 1.1cm 0.25cm 0.25cm, clip=true]{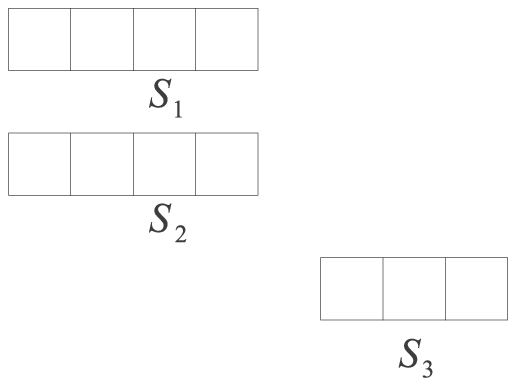}
\label{fig:num_tuples_canonical2}
}
\subfigure[] {
\includegraphics[trim=0.25cm 1.1cm 0.25cm 0.5cm, clip=true]{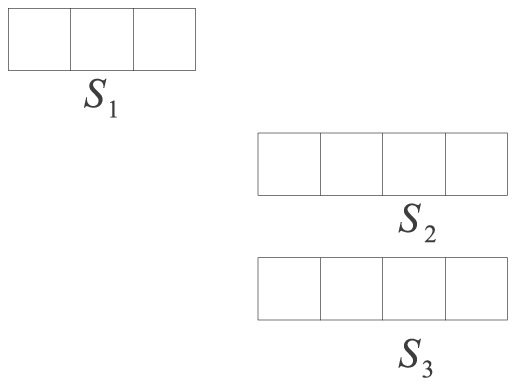}
\label{fig:num_tuples_canonical3}
}
\hspace{-3em}
\subfigure[] {
\includegraphics[trim=0.25cm 1.1cm 0.25cm 0.5cm, clip=true]{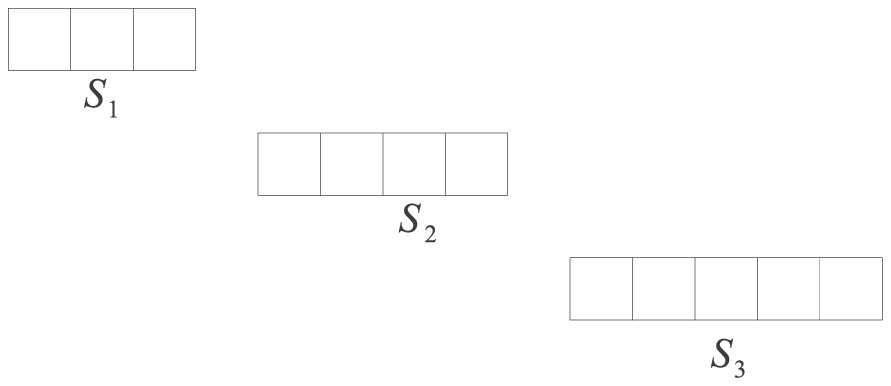}
\label{fig:num_tuples_canonical4}
}
\caption{For $n = 3$, four canonical cases of the three ``valid''
  (i.e. the particle indices in each node are in increasing
  depth-first order) node tuples encountered during the algorithm: (a)
  All three nodes are equal; (b) $S_1$ and $S_2$ are equal, and $S_3$
  comes later in the depth-first order; (c) $S_2$ and $S_3$ are equal
  and come later in the depth-first order; (d) All three nodes are
  different.
\label{fig:num_tuples}
}
\end{figure}

We use a variant of $kd$-trees~\cite{bentley1975multidimensional} to
form hierarchical groupings of points based on their locations using
the recursive procedure shown in
Algorithm~\ref{alg:multibody:buildkdtree}.  Initially, the algorithm
starts with $P = X$ (the entire point set).  We split a given set of
points along the widest dimension of the bounding hyper-rectangle into
two equal halves at the splitting coordinate. We continue splitting
until the number of points is below some user-defined threshold called
the {\it leaf threshold}. If the number of points owned by a node
exceeds the leaf threshold, then it is called an {\it internal
  node}. Otherwise it is called a {\it leaf node}. Assuming that each
split on a level results in the equal number of points on the left
subset and the right subset $P^L$ and $P^R$ respectively, the runtime
cost is $\mathcal{O}( | X| \log | X | )$. We note that the cost of
building a $kd$-tree is negligible compared to the actual multibody
potential computation (see
Section~\ref{sec:experimental_results}). See
Figure~\ref{fig:multibody:kdtree_dfs_index}.
{\small
\begin{algorithm}

\caption{$\mbox{\buildkdtree}(P)$\label{alg:multibody:buildkdtree}}

\begin{algorithmic}

\IF{ $| P |$  is above the leaf threshold}

\STATE{Find the widest dimension $d$ of the bounding box of $P$.}

\STATE{Choose an axis-aligned split $s$ along $d$.}

\STATE{Split $P = P^L \cup P^R$ where $P^L = \{ x \in P \ | \ x[d]
  \leq s \}$ and $P^R = P \backslash P^L $.}

\STATE{$\mbox{\buildkdtree}(P^L)$, $\mbox{\buildkdtree}(P^R)$}

\STATE{Form far-field moments of $P$ by translating far-field moments
  of $P^L$ and $P^R$.}

\ELSE

\STATE{Form far-field moments of $P$.}

\ENDIF

\STATE{Initialize summary statistics of $P$.}

\end{algorithmic}

\end{algorithm}
}
The general framework for computing
Equation~\eqref{eq:canonical_multibody_summation} is formalized
in~\cite{gray2001nbp,gray2003nde,gray2003vfm,gray2003rem}. This
approach consists of the following steps:
\begin{enumerate}
\item{Build a spatial tree (such as {\it kd}-trees) for the set of
  particles $X$ and build far-field moments on each node of the tree
  {\bf (Bottom-up phase)}.}
\item{Perform a {\it multi-tree traversal} over
$n$-tuples of nodes {\bf (Approximation phase)}.}
\item{Pre-order traverse the tree and propagate unincorporated bound
  changes downward {\bf (Top-down phase)}.}
\end{enumerate}
\begin{figure}[t]
\centering
\scalebox{0.425}{\includegraphics[trim=0em 0em 0em 0em, clip=true]{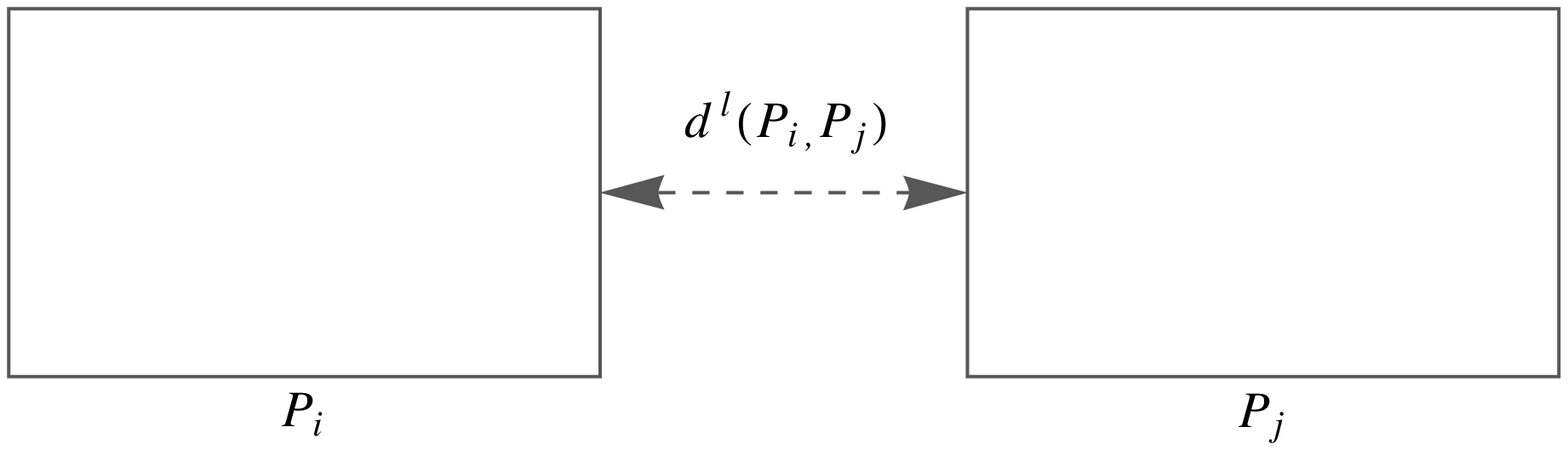}}
\scalebox{0.425}{\includegraphics[trim=0em 0em 0em 0em, clip=true]{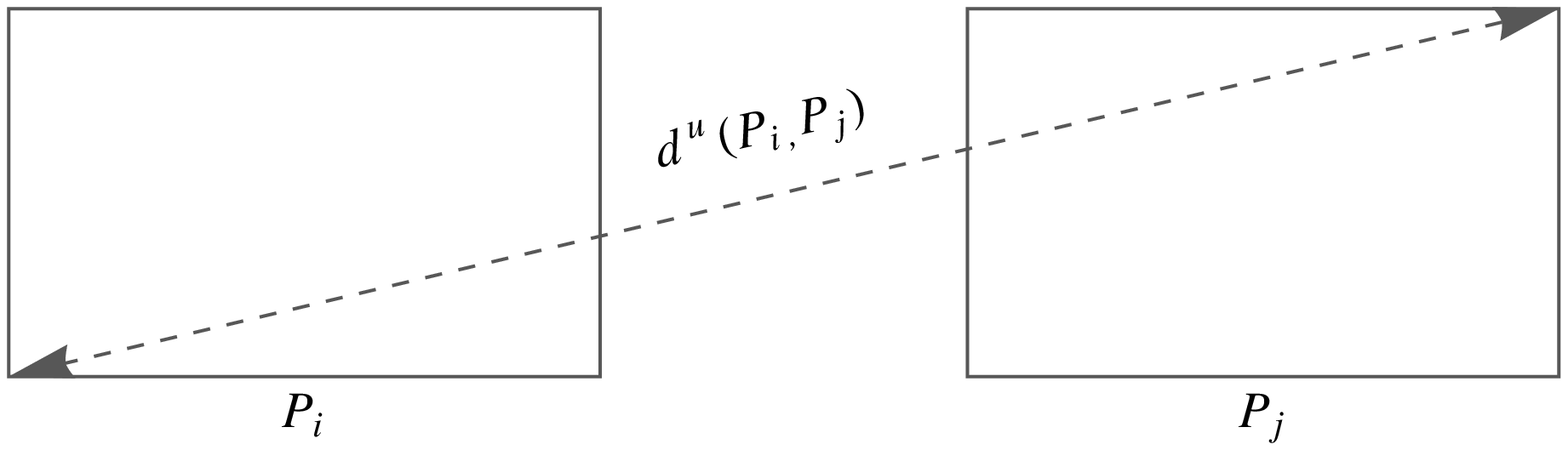}}
\caption{The lower and upper bound on pairwise distances between the
points contained in a pair of nodes.}
\label{fig:distance_bounds}
\end{figure}
Step 2 utilizes the procedure shown in
Algorithm~\ref{alg:multibody:mtpotentialmain} (called by setting each
$P_i = X$ for $1 \leq i \leq n$), a recursive function that allows us
to consider the $n$-tuples formed by choosing each $x_i$ from $P_i$;
we can gain efficiency over the naive enumeration of the $n$-tuples by
using the bounding box and the moment information stored in each
$P_i$. One such information is the distance bound computed using the
bounding box (see
Figure~\ref{fig:distance_bounds}).
{\small
\begin{algorithm}[t]
\caption{$\mbox{\mtpotentialcanonical}(\{P_i\}_{i=1}^n )$\label{alg:multibody:mtpotentialmain}}

\begin{algorithmic}

\IF {$\mbox{\cansummarize}( \{P_i\}_{i=1}^n )$ \ \ \ (Try
  approximation.)}

\STATE{$\mbox{\summarize}(\{P_i\}_{i=1}^n, \epsilon, \tau, \alpha )$}

\ELSE

\IF{all of $S_i$ are leaves}

\STATE{$\mbox{\mtpotentialbase}(\{P_i\}_{i=1}^n )$ \ \ \ (Base case.)}

\ELSE

\STATE{Find an internal node $P_k$ to split among $\{ P_i \}_{i = 1}^n$.}

\STATE{Propagate bounds of $P_k$ to $P_k^L$ and $P_k^R$.}

\STATE{$\mbox{\mtpotentialcanonical}(\{P_1, \cdots, P_{k - 1}, P_k^L, P_{k + 1}, \cdots, P_n \}  )$}

\STATE{$\mbox{\mtpotentialcanonical}(\{P_1, \cdots, P_{k - 1}, P_k^R, P_{k + 1}, \cdots, P_n  \}  )$}

\STATE{Refine summary statistics based on the two recursive calls.}

\ENDIF

\ENDIF

\end{algorithmic}

\end{algorithm}
}
$\mbox{\cansummarize}$ function first eliminates redundant recursive
calls for the list of node tuples that satisfy the following
condition: if there exists a pair of nodes $P_i$ and $P_j$ ($i < j$)
among the node list $P_1, \cdots, P_n$, such that the maximum
depth-first rank of $P_i$ is less than the minimum depth-first rank of
$P_j$. In this case, the function returns true. See
Figure~\ref{fig:multibody:kdtree_dfs_index} and~\cite{moore2001fast}.
In addition, if any one of the nodes in the list includes one of the
other nodes (i.e. there exists nodes $P_i$ and $P_j$ such that the
minimum depth-first rank of $P_i <$ the minimum depth-first rank of
$P_j <$ the maximum depth-first rank of $P_j <$ the maximum
depth-first rank of $P_i$), $\mbox{\cansummarize}$ returns false. We
do this because it is a bit tricky to count the number of tuples for
each point in this case (see Figure~\ref{fig:num_tuples}).

Otherwise, $\mbox{\cansummarize}$ function tests whether each
potential sum for $x \in \bigcup\limits_{1 \leq m \leq n} P_m$ can be
approximated within the error tolerance determined by the algorithm.
For example, if $n = 4$, we test for each $x_1 \in P_1$, $x_2 \in
P_2$, $x_3 \in P_3$, $x_4 \in P_4$, the following exact quantities can
be approximated:
{\small
\begin{align*}
\Phi(x_1; P_2 \times P_3 \times P_4 ) &= \sum\limits_{ x_{i_2}
  \in P_2 \backslash \{ x_1 \} } \ 
\sum\limits_{ \substack{ x_{i_3} \in P_3 \backslash \{ x_1 \}  \\ i_2 < i_3 } } \
\sum\limits_{ \substack{x_{i_4} \in P_4 \backslash \{ x_1 \} \\ i_{3} < i_4 } }
\phi ( x_1, x_{i_2}, x_{i_3}, x_{i_4} )\\
\Phi(x_2; P_1 \times P_3 \times P_4 ) &= \sum\limits_{ x_{i_1}
  \in P_1 \backslash \{ x_2 \} } \ 
\sum\limits_{ \substack{ x_{i_3} \in P_3 \backslash \{ x_2 \}  \\ i_1 < i_3 } } \
\sum\limits_{ \substack{x_{i_4} \in P_4 \backslash \{ x_2 \} \\ i_{3} < i_4 } }
\phi ( x_2, x_{i_1}, x_{i_3}, x_{i_4} )
\end{align*}
\begin{align*}
\Phi(x_3; P_1 \times P_2 \times P_4 ) &= \sum\limits_{ x_{i_1}
  \in P_1 \backslash \{ x_3 \} } \ 
\sum\limits_{ \substack{ x_{i_2} \in P_2 \backslash \{ x_3 \}  \\ i_1 < i_2 } } \
\sum\limits_{ \substack{x_{i_4} \in P_4 \backslash \{ x_3 \} \\ i_{2} < i_4 } }
\phi ( x_3, x_{i_1}, x_{i_2}, x_{i_4} )\\
\Phi(x_4; P_1 \times P_2 \times P_3 ) &= \sum\limits_{ x_{i_1}
  \in P_1 \backslash \{ x_4 \} } \ 
\sum\limits_{ \substack{ x_{i_2} \in P_2 \backslash \{ x_4 \}  \\ i_1 < i_2 } } \
\sum\limits_{ \substack{x_{i_3} \in P_3 \backslash \{ x_4 \} \\ i_2 < i_3 } }
\phi ( x_4, x_{i_1}, x_{i_2}, x_{i_3} )
\end{align*}
}If the approximation is not possible, then the
algorithm continues to consider the data at a finer granularity; it
chooses an internal node $P_k$ (typically the one with the largest
diameter) to split among $\{ P_i \}_{i = 1}^n$. Before recursing to
two sub-calls in Line 9 and Line 10 of
Algorithm~\ref{alg:multibody:mtpotentialmain}, the algorithm can
optionally push quantities from a node that is being split to its
child nodes (Line 8). After returning from the recursive calls, the
node that was just split can refine {\it summary statistics} based on
the results accumulated on its child nodes. The details of these
operations are available in earlier
papers~\cite{gray2001nbp,gray2003nde,gray2003vfm,gray2003rem,1102.2878}.

The basic idea is to terminate the recursion as soon as possible,
i.e. by considering a tuple of large subsets and avoiding the number
of exhaustive leaf-leaf-leaf computations. We note that the
$\mbox{\cansummarize}$ and $\mbox{\summarize}$ functions effectively
replace unwieldy interaction lists used in FMM algorithms. Interaction
lists in $n$-tuple interaction, if naively enumerated, can be large
depending on the potential function $\phi$ and the dimensionality $D$
of the problem, whereas the generalized $N$-body approach can handle a
wide spectrum of problems without this drawback.

\begin{figure}[t]
\centering
\scalebox{0.5}{\includegraphics[trim=0em 0em 0em 0em,clip=true]{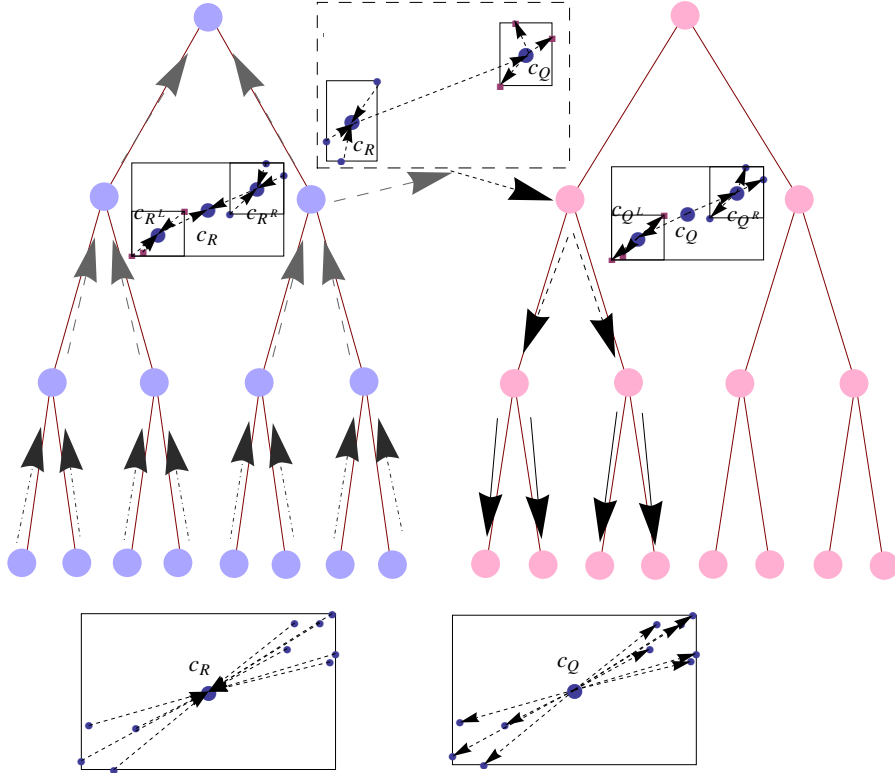}}
\caption{The reference points (the left tree) are hierarchically
  compressed and uncompressed when a pair of query (from the right
  tree)/reference nodes is approximated within an error
  tolerance.\label{fig:sdm2012:fmm_plot}}
\end{figure}

\subsection{Algorithm for Pairwise Potentials ($n = 2$)}
\label{sec:algorithm_for_pairwise_potentials}
The general algorithmic strategy for pairwise potentials $\phi(\cdot,
\cdot)$ is described
in~\cite{gray2001nbp,gray2003nde,gray2003vfm,gray2003rem}, and
consists of the following three main phases (see Figure~\ref{fig:sdm2012:fmm_plot}). 
Suppose we are given a
set of ``source'' points (denoted as {\it reference points}) and a set
of ``target'' points (denoted as {\it query points})\footnote{The terms ``reference/query'' have been used in the general framework we are applying}.
\begin{enumerate}
\item{{\bf Bottom-up phase: }Compute far-field moments of order $p$ in
every leaf node of the reference tree. The resulting far-field
expansion of each reference node $P_2$ is given by:
{ \small
\begin{align}
\Phi(x; P_2 ) = \sum\limits_{ \boldsymbol\alpha \geq 0} \left [
  \sum\limits_{x_{i_2} \in P_2} \frac{(-1)^{\boldsymbol\alpha}}{\boldsymbol\alpha !} (
  x_{i_2} - c_{P_2} )^{\boldsymbol\alpha} \right ] D^{\boldsymbol\alpha} \phi (x - c_{P_2} )
\notag = \sum\limits_{ \boldsymbol\alpha \geq 0 } M_{\boldsymbol\alpha}(P_2, c_{P_2} )
D^{ \boldsymbol\alpha} \phi ( x - c_{P_2} )
\label{eq:two_body_farfield_expansion}
\end{align}
}$\Phi(x; P_2)$ reads as ``the potential sum on $x$ due to the
contribution of $P_2$" and $M_{\boldsymbol\alpha}(P_2, c_{P_2} )$ as ``the
$\boldsymbol\alpha$-th far-field coefficient of $P_2$ centered at $c_{P_2}$."
Because it is impossible to store an infinite number of far-field
moments $M_{\boldsymbol\alpha}(P_2, c_{P_2} )$, we truncate the Taylor expansion
up to the order $p$ (determined either arbitrarily or by an
appropriate error criterion): { \small
\begin{align}
\widetilde{ \Phi}(x; P_2; F(c_{P_2}, p) ) = \sum\limits_{ | \boldsymbol\alpha | \leq p }
M_{\boldsymbol\alpha}(P_2, c_{P_2} ) D^{ \boldsymbol\alpha} \phi (x - c_{P_2} )
\end{align}
}such that $ \left | \widetilde{ \Phi}(x ; P_2) - \Phi(x; P_2) \right
|$ is sufficiently small. $\widetilde{ \Phi}(x; P_2; F(c_{P_2}, p) )$
reads as ``the approximated potential sum on $x$ due to the points
owned by $P_2$ using up to the $p$-th order far-field expansion of
$P_2$ centered at $c_{P_2}$."

For internal reference nodes, perform the far-to-far (F2F) translation
to convert the far-field moments owned by the child nodes to form the
far-field moments for their common parent node $P_2$. For example, the
far-field moments of $P^L_2$ centered at $c_{P^L_2}$ is shifted to
$c_{P_2}$ by:{ \small
\begin{align}
\widetilde{ \Phi }(x; P_2^L; F(c_{P_2}, p) ) = \sum\limits_{ \boldsymbol\gamma
  \leq p} M_{\boldsymbol\gamma}(P_2^L, c_{P_2}) (-1)^{ \boldsymbol\gamma } D^{ \boldsymbol\gamma } \phi
( x - c_{P_2} )
\label{eq:f2f_pairwise}
\end{align}
} where 
{\small
\begin{equation}
M_{\boldsymbol\gamma}(P_2^L, c_{P_2}) = \sum\limits_{ \boldsymbol\alpha \leq \boldsymbol\gamma }
\frac{ M_{\boldsymbol\alpha}( P_2^L, c_{P_2^L}) ( c_{P_2^L} - c_{P_2} )^{ \boldsymbol\gamma - \boldsymbol\alpha
} } { ( \boldsymbol\gamma - \boldsymbol\alpha ) !}
\label{eq:two_body_f2f_translation}
\end{equation}
 }Note that there is no error incurred in each F2F translation, i.e.
$\widetilde{ \Phi }(x; P_2^L; F(c_{P_2^L}, p) ) = \widetilde{ \Phi
}(x; P_2^L; F(c_{P_2}, p) )$ for any query point $y$ from the
intersection of the domains of $x$ for $\widetilde{ \Phi }(x; P_2^L;
F(c_{P_2^L}, p) ) $ and $\widetilde{ \Phi }(x; P_2^L; F(c_{P_2}, p)
)$; the domain for which the far-field expansion remains valid depends
on the error bound criterion for each potential. The far-field moments
of the parent node $P_2$ is the sum of the translated moments of its
child nodes: $M_{\boldsymbol\gamma}(P_2, c_{P_2}) = \sum\limits_{ \boldsymbol\alpha \leq
  \boldsymbol\gamma} \frac{ M_{\boldsymbol\alpha}(P_2^L, c_{P_2^L} ) ( c_{P_2^L} - c_{P_2}
  )^{ \boldsymbol\gamma - \boldsymbol\alpha } } { ( \boldsymbol\gamma - \boldsymbol\alpha ) !} + \frac{ M_{ \boldsymbol\alpha
  }(P_2^R, c_{P_2^R} ) ( c_{P_2^R} - c_{P_2} )^{ \boldsymbol\gamma - \boldsymbol\alpha } } {
  ( \boldsymbol\gamma - \boldsymbol\alpha ) !}  $ }
\item{{\bf Approximation phase: }For a given pair of the query and the
  reference nodes, determine the order of approximation and either (1)
  translate the far-field moments of the reference node to the local
  moments of the query node (2) or recurse to their subsets, if the
  F2L translation is more costly than the direct exhaustive method.

Let us re-write the exact contribution of $P_2$ to a point $x \in
P_1$: { \small
\begin{align}
& \Phi(x ; P_2) = \sum\limits_{ \boldsymbol\beta \geq 0} \frac{1}{\boldsymbol\beta !}
  \sum\limits_{ \boldsymbol\alpha \geq 0} M_{\boldsymbol\alpha}(P_2, c_{P_2}) D^{ \boldsymbol\alpha +
    \boldsymbol\beta } \phi ( c_{P_1} - c_{P_2} ) ( x - c_{P_1} )^{ \boldsymbol\beta }
  \notag \\ =& \sum\limits_{ \boldsymbol\beta \geq 0 } \left [ \sum\limits_{
      x_{i_2} \in P_2 } \frac{1}{ \boldsymbol\beta !} D^{ \boldsymbol\beta} \phi ( c_{P_1} -
    x_{i_2}) \right ] ( x - c_{P_1} )^{ \boldsymbol\beta } = \sum\limits_{ \boldsymbol\beta
    \geq 0} N_{\boldsymbol\beta}(P_2, c_{P_1} ) (x - c_{P_1} )^{ \boldsymbol\beta }
\label{eq:direct_local_pairwise}
\end{align}
}where $N_{\boldsymbol\beta}(P_2, c_{P_1})$ reads as ``the exact local moments
\footnote{We use $N$ to denote the local moments because a
  ``near-field'' expansion is another widely used term for a local
  expansion. It avoids the potential notational confusion in the later
  parts of the paper.}  contributed by the points in $P_2$ centered at
$c_{P_1}$.'' Truncating Equation~\eqref{eq:direct_local_pairwise} at
$| \boldsymbol\beta | \leq p'$ for some $p' \leq p$ yields a direct local
accumulation of order $p$.

From the bottom-up phase, we know that $| \boldsymbol\alpha | \leq p$. Similarly,
we can store only a finite number of local moments up to the order $p'
\leq p$ and thus $| \boldsymbol\beta | \leq p'$. We get the local expansion for
$P_1$ formed due to translated far-field moments of $P_2$: { \small
\begin{align}
\widetilde{ \Phi} (x ; P_2 ; \widetilde{N}(c_{P_1}, p') ) &=
\sum\limits_{ | \boldsymbol\beta | \leq p' } \left [ \frac{1}{ \boldsymbol\beta !}
  \sum\limits_{ | \boldsymbol\alpha | \leq p' } M_{\boldsymbol\alpha}(P_2, c_{P_2} ) D^{
    \boldsymbol\alpha + \boldsymbol\beta } \phi_{1, 2} ( c_{P_1} - c_{P_2} ) \right ] ( x -
c_{P_1} )^{ \boldsymbol\beta } \notag \\ &= \sum\limits_{ | \boldsymbol\beta | \leq p' }
\widetilde{N}_{\boldsymbol\beta}(P_2, c_{P_1} ) ( x - c_{P_1} )^{ \boldsymbol\beta}
\label{eq:two_body_f2l_translation}
\end{align}
}where $\widetilde{N}_{\boldsymbol\beta}( P_2, c_{P_1} )$ reads as
``approximation to the exact local moments $ N_{\boldsymbol\beta}( P_2, c_{P_1} )
$" and $\widetilde{\Phi}( x ; P_2 ; \widetilde{N}( c_{P_1}, p') )$ as
``the approximated potential sum on $x$ due to the points in $P_2$
using up to the $p$-th order inexact local moments centered at
$c_{P_1}$". The F2L translation is applied only if $\left |
\widetilde{\Phi}( x ; P_2 ; \widetilde{N}(c_{P_1}, p')) - \Phi( x; P_2
) \right |$ is sufficiently small.}
\item{{\bf Top-down phase: }Propagate the local moments of each query
  node (i.e. pruned quantities) to its child nodes using the
  local-to-local (L2L) operator. Suppose we have the following local
  expansion for $x \in P_1$: { \small
$$\widetilde{\Phi}(x; \mathit{F2L}(P_1) \cup \mathit{DL}(P_1) ;
    \widetilde{N}( c_{P_1}, p_{P_1}^u)) = \sum\limits_{ | \boldsymbol\alpha |
      \leq p_{P_1}^{u}} \widetilde{N}_{\boldsymbol\alpha}( \mathit{F2L}(P_1) \cup
    \mathit{DL}(P_1), c_{P_1} ) (x_{i_1} - c_{P_1} )^{ \boldsymbol\alpha }
 $$}where $p_{P_1}^u$ is the maximum approximation order among (1) the
  F2L translations performed for $P_1$ and all of the ancestor nodes
  of $P_1$ (denoted by $\mathit{F2L}(P_1)$); and (2) the direct local
  accumulations of $P_1$ and those passed down from all of the
  ancestors of $P_1$ (denoted by $\mathit{DL}(P_1)$). Shifting the
  expansion to another center $c_{P_1}^* \in P_1$ is given by: {
    \small
\begin{align}
& \widetilde{\Phi}(x ; \mathit{F2L}(P_1) \cup \mathit{DL}(P_1);
\widetilde{N}( c_{P_1}^*, p_{P_1}^u) ) \\ =& \sum\limits_{ | \boldsymbol\alpha |
  \leq p_{P_1}^u } \left [ \sum\limits_{ \boldsymbol\beta \geq \boldsymbol\alpha}
  \binom{\boldsymbol\beta}{\boldsymbol\alpha} \widetilde{N}_{\boldsymbol\beta}( \mathit{F2L}(P_1) \cup
  \mathit{DL}(P_1), c_{P_1}) ( c_{P_1}^* - c_{P_1} )^{ \boldsymbol\beta - \boldsymbol\alpha
  } \right ] (x - c_{P_1}^* )^{ \boldsymbol\alpha } \notag \\ =& \sum\limits_{ |
  \boldsymbol\alpha | \leq p_{P_1}^u } \widetilde{N}_{\boldsymbol\alpha} ( \mathit{F2L}(P_1)
\cup \mathit{DL}(P_1), c_{P_1}^* ) ( x - c_{P_1}^* )^{ \boldsymbol\alpha }
\label{eq:two_body_l2l_translation}
\end{align}
}This shifted moments are added to the local moments of each child of
  $P_1$, in effect transmitting the pruned contributions downward. At
  each query leaf, we evaluate the resulting local expansion at each
  query point.  }
\end{enumerate}

\begin{figure}[t]
\centering
\scalebox{0.75}{ \includegraphics{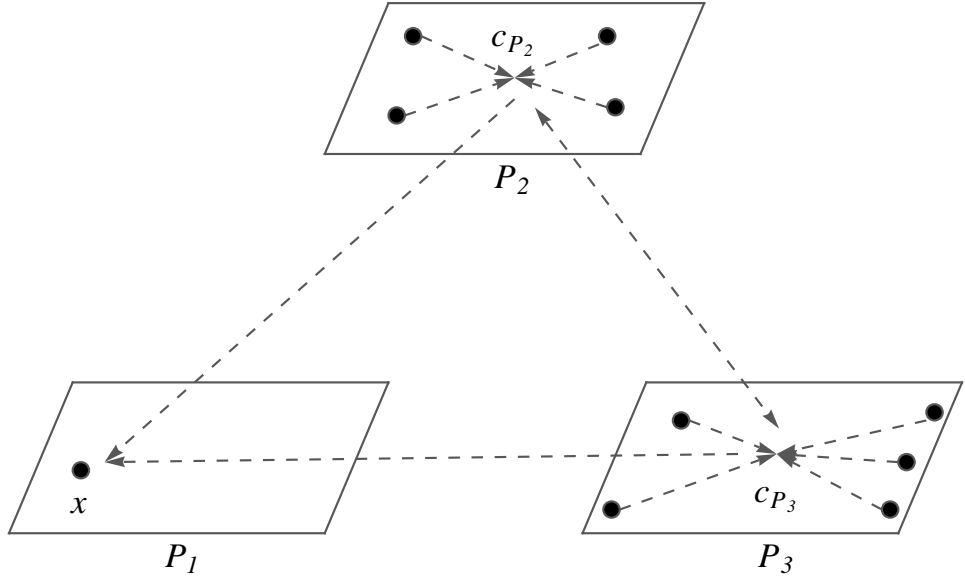}}
\caption{A far-field expansion at $x_{i_1}$ created by the moments of
  $P_2$ and $P_3$. Note the double-arrow between the nodes $P_2$ and
  $P_3$ corresponding to the basis functions $D^{\boldsymbol\alpha - \boldsymbol\alpha_{1,
      2} - \boldsymbol\alpha_{1, 3} } \phi_{2, 3}(P_2^c - P_3^c)$ (see
  Equation~\ref{eq:multibody:three_body_farfield_expansion}).
\label{fig:multibody:three_body_farfield_expansion}
}
\end{figure}

\subsection{Far-field Expansion for Three-body Potentials ($n = 3$)}
In this section, we define far-field expansions for a three-body
potential that is a product of functions of pairwise distances (see
Equation~\eqref{eq:canonical_multibody_summation}): 
{\small
\begin{equation}
\phi(x_{i_1},
x_{i_2}, x_{i_3}) = \phi_{1,2}( x_{i_1}, x_{i_2} ) \cdot \phi_{1,3}(
x_{i_1}, x_{i_3} ) \cdot \phi_{2,3}( x_{i_2}, x_{i_3} )
\label{eq:three_body_potential}
\end{equation}
}
We define the far-field moments of a node
  the same way defined for the pairwise potential case. Suppose we are
  given three nodes $P_1 \not = P_2 \not = P_3$ from the tree. The
  following $(n - 1)$-nested sum expresses the contribution for $x \in
  P_1$ due to the other nodes $P_2$ and $P_3$: { \small
\begin{equation}
 \Phi(x ; P_2 \times P_3 ) = \sum\limits_{ x_{i_2} \in P_2 }
 \sum\limits_{ x_{i_3} \in P_3 } \phi(x, x_{i_2}, x_{i_3})
\label{eq:triple_subset_summation}
\end{equation} }
The basic goal here is to decompose
Equation~\eqref{eq:triple_subset_summation} into sums of products of
the far-field moments of each node. A far-field expansion for $x_{i_1}
\in P_1$ induced by the far-field moments of $P_2$ and $P_3$ is given
by (see Figure~\ref{fig:multibody:three_body_farfield_expansion}):
{\small
\begin{align*}
& \Phi(x ; P_2 \times P_3 ) \\ = &
 \sum\limits_{ x_{i_2} \in P_2 }
 \sum\limits_{ x_{i_3} \in P_3 }
\sum\limits_{ \boldsymbol\alpha_{1,2} \geq 0 } \frac{( x_{i_2} - c_{P_2})^{\boldsymbol\alpha_{1,2}}}{\boldsymbol\alpha_{1,2}!}  (-1)^{ \boldsymbol\alpha_{1, 2} } D^{\boldsymbol\alpha_{1,2}} \phi_{1,2} ( x - c_{P_2} ) \\
& \sum\limits_{ \boldsymbol\alpha_{1,3} \geq 0 } \frac{( x_{i_3} - c_{P_3})^{\boldsymbol\alpha_{1,3}}}{\boldsymbol\alpha_{1,3}!} (-1)^{ \boldsymbol\alpha_{1,3} }D^{\boldsymbol\alpha_{1,3}} \phi_{1,3} ( x - c_{P_3} ) \\
& \sum\limits_{ \boldsymbol\alpha_{2, 3} \geq 0 }
\sum\limits_{ \boldsymbol\beta_{2, 3} \leq \boldsymbol\alpha_{2, 3} }
\frac{ ( x_{i_2} - c_{P_2} )^{ \boldsymbol\beta_{2, 3} }}{ \boldsymbol\beta_{2, 3} ! } \frac{ ( x_{i_3} - c_{P_3} )^{ \boldsymbol\alpha_{2,3} - \boldsymbol\beta_{2,3} } }{ ( \boldsymbol\alpha_{2,3} - \boldsymbol\beta_{2,3} ) ! } (-1)^{ \boldsymbol\alpha_{2,3} - \boldsymbol\beta_{2,3} } D^{ \boldsymbol\alpha_{2, 3} } \phi_{2,3} ( c_{P_2} - c_{P_3} )
\end{align*}
}By setting $\boldsymbol\alpha =
\boldsymbol\alpha_{1,2} + \boldsymbol\alpha_{1,3} +
\boldsymbol\alpha_{2,3}$ and pushing the summations over $x_{i_2} \in
P_2$ and $x_{i_3} \in P_3$ inside, we get: {\small
\begin{align}
 \Phi(x; P_2 \times P_3 ) =& \sum\limits_{\boldsymbol\alpha \geq 0} \sum\limits_{
   \boldsymbol\alpha_{1, 2} \leq \boldsymbol\alpha } \sum\limits_{ \boldsymbol\alpha_{1, 3} \leq \boldsymbol\alpha
   - \boldsymbol\alpha_{1, 2} } \sum\limits_{ \boldsymbol\beta_{2, 3} \leq \boldsymbol\alpha -
   \boldsymbol\alpha_{1, 2} - \boldsymbol\alpha_{1, 3} } \binom{\boldsymbol\alpha_{1,2} + \boldsymbol\beta_{2,
     3}}{\boldsymbol\alpha_{1,2}} \binom{ \boldsymbol\alpha - \boldsymbol\alpha_{1,2} - \boldsymbol\beta_{2, 3}}{
   \boldsymbol\alpha_{1,3}} \notag \\ & M_{\boldsymbol\alpha_{1,2} + \boldsymbol\beta_{2,3}}( P_2,
 c_{P_2}) M_{ \boldsymbol\alpha - \boldsymbol\alpha_{1,2} - \boldsymbol\beta_{2,3} }(P_3, c_{P_3} )
 (-1)^{\boldsymbol\beta_{2,3}} \notag \\ & D^{\boldsymbol\alpha_{1, 2} } \phi_{1,
   2} ( x_{i_1} - c_{P_2} ) D^{ \boldsymbol\alpha_{1, 3} } \phi_{1, 3} ( x_{i_1}
 - c_{P_3} ) D^{ \boldsymbol\alpha - \boldsymbol\alpha_{1, 2} - \boldsymbol\alpha_{1, 3} } \phi_{2, 3}
 ( c_{P_2} - c_{P_3} )
\label{eq:multibody:three_body_farfield_expansion}
\end{align}
}Truncating $\boldsymbol \alpha$ at $p$-th order yields:
{\small
\begin{align}
&  \widetilde{\Phi}(x; P_2 \times P_3; F(  c_{P_2} \times c_{P_3} , p) ) \notag \\
=& \sum\limits_{ | \boldsymbol\alpha | \leq p} \sum\limits_{
   \boldsymbol\alpha_{1, 2} \leq \boldsymbol\alpha } \sum\limits_{ \boldsymbol\alpha_{1, 3} \leq \boldsymbol\alpha
   - \boldsymbol\alpha_{1, 2} } \sum\limits_{ \boldsymbol\beta_{2, 3} \leq \boldsymbol\alpha -
   \boldsymbol\alpha_{1, 2} - \boldsymbol\alpha_{1, 3} } \binom{\boldsymbol\alpha_{1,2} + \boldsymbol\beta_{2,
     3}}{\boldsymbol\alpha_{1,2}} \binom{ \boldsymbol\alpha - \boldsymbol\alpha_{1,2} - \boldsymbol\beta_{2, 3}}{
   \boldsymbol\alpha_{1,3}} \notag \\ & M_{\boldsymbol\alpha_{1,2} + \boldsymbol\beta_{2,3}}( P_2,
 c_{P_2}) M_{ \boldsymbol\alpha - \boldsymbol\alpha_{1,2} - \boldsymbol\beta_{2,3} }(P_3, c_{P_3} )
 (-1)^{\boldsymbol\beta_{2,3}} \notag \\ & D^{\boldsymbol\alpha_{1, 2} } \phi_{1,
   2} ( x_{i_1} - c_{P_2} ) D^{ \boldsymbol\alpha_{1, 3} } \phi_{1, 3} ( x_{i_1}
 - c_{P_3} ) D^{ \boldsymbol\alpha - \boldsymbol\alpha_{1, 2} - \boldsymbol\alpha_{1, 3} } \phi_{2, 3}
 ( c_{P_2} - c_{P_3} )
\label{eq:multibody:truncated_three_body_farfield_expansion}
\end{align}
}where $\widetilde{\Phi}(x; P_2 \times P_3; F( c_{P_2} \times c_{P_3}
, p) )$ reads as ``the $p$-th order far-field expansion at $x$ due to
the moments of $P_2$ centered at $c_{P_2}$ and the moments of $P_3$
centered at $c_{P_3}$.'' 

{\noindent \bf Computational Cost of Evaluating the Far-field
  Expansion. }The first three summations over $\boldsymbol\alpha$,
$\boldsymbol \alpha_{1,2}$, $\boldsymbol \alpha_{1,3}$ collectively
contribute $\mathcal{O}(p^3)$ terms, and the inner summation
contributing at most $\mathcal{O}(p^3)$ terms. Thus, evaluating the
$p$-th order far-field expansion for a three-body potential on a
single point takes $\mathcal{O} \left ( p^6 \right )$ time.

\subsection{Far-field Expansion for General Multibody Potentials $(n \geq 2)$}
For a general multibody potential that can be expressed as products of
pairwise functions (see Equation~\eqref{eq:multibody_potential_form}),
the far-field expansion induced by the points in $P_2, \cdots, P_n$
for $x \in P_1$ is: {\small
\begin{align*}
& \Phi(x ; P_2 \times \cdots \times P_n ) \\ = &
\prod\limits_{2 \leq k \leq n}
 \sum\limits_{ x_{i_k} \in P_k }
\sum\limits_{ \boldsymbol\alpha_{1,k} \geq 0 } \frac{( x_{i_k} - c_{P_k})^{\boldsymbol\alpha_{1,k}}}{\boldsymbol\alpha_{1,k}!}  (-1)^{ \boldsymbol\alpha_{1, k} } D^{\boldsymbol\alpha_{1,k}} \phi_{1,k} ( x - c_{P_k} ) \\
& 
\prod\limits_{2 \leq s < t \leq n}
\sum\limits_{ \boldsymbol\alpha_{s, t} \geq 0 }
\sum\limits_{ \boldsymbol\beta_{s, t} \leq \boldsymbol\alpha_{s, t} }
\frac{ ( x_{i_s} - c_{P_s} )^{ \boldsymbol\beta_{s, t} }}{ \boldsymbol\beta_{s, t} ! } \frac{ ( x_{i_t} - c_{P_t} )^{ \boldsymbol\alpha_{s,t} - \boldsymbol\beta_{s,t} } }{ ( \boldsymbol\alpha_{s,t} - \boldsymbol\beta_{s,t} ) ! } (-1)^{ \boldsymbol\alpha_{s,t} - \boldsymbol\beta_{s,t} } D^{ \boldsymbol\alpha_{s, t} } \phi_{s,t} ( c_{P_s} - c_{P_t} )
\end{align*}
}Focus on grouping and multiplying monomial powers of $(x_{i_k} -
c_{P_k})$ for each $2 \leq k \leq n$: {\small
\begin{align*}
\frac{ (x_{i_k} - c_{P_k})^{ \boldsymbol\alpha_{1,k} + \sum\limits_{u = 2}^{k - 1} (\boldsymbol\alpha_{u,k} - \boldsymbol\beta_{u,k}) + \sum\limits_{v = k + 1}^n \boldsymbol\beta_{k, v} } }{  \boldsymbol\alpha_{1,k} ! \prod\limits_{u = 2}^{k - 1} (\boldsymbol\alpha_{u,k} - \boldsymbol\beta_{u,k}) ! \prod\limits_{v = k + 1}^n \boldsymbol\beta_{k, v}  !  }
\end{align*}
}Let $\boldsymbol\xi_{k} = \boldsymbol\alpha_{1,k} + \sum\limits_{u =
  2}^{k - 1} (\boldsymbol\alpha_{u,k} - \boldsymbol\beta_{u,k}) +
\sum\limits_{v = k + 1}^n \boldsymbol\beta_{k, v} $ and $b_k = \frac{
  \boldsymbol\xi_k ! }{\boldsymbol\alpha_{1,k} ! \prod\limits_{u = 2}^{k
    - 1} (\boldsymbol\alpha_{u,k} - \boldsymbol\beta_{u,k}) !
  \prod\limits_{v = k + 1}^n \boldsymbol\beta_{k, v} !} $. Then,
           {\small
\begin{align}
& \Phi(x ; P_2 \times \cdots \times P_n ) \notag \\ 
= & 
\prod\limits_{2 \leq s < t \leq n}
\prod\limits_{2 \leq
    k \leq n} 
\sum\limits_{  \boldsymbol \alpha_{1, k} \geq 0 }
\sum\limits_{ \boldsymbol\alpha_{s, t} \geq 0 }
\sum\limits_{ \boldsymbol\beta_{s, t} \leq \boldsymbol\alpha_{s, t} }
b_k \ M_{ \boldsymbol\xi_k }(P_k, c_{P_k}) \ (-1)^{ \boldsymbol\beta_{s,t}} \ D^{\boldsymbol\alpha_{1,k}}
  \phi_{1,k} ( x - c_{P_k} )
 D^{ \boldsymbol\alpha_{s, t} } \phi_{s,t} ( c_{P_s} - c_{P_t} )
\label{eq:multibody:multibody_farfield_expansion}
\end{align}
}Equation~\eqref{eq:multibody:multibody_farfield_expansion} is a
           convolution of far-field moments of $P_2, \cdots, P_n$. We
           can truncate the expansion above for terms for $|
           \boldsymbol\alpha | = \left | \sum\limits_{1 \leq r < s
             \leq n} \alpha_{r, s} \right | > p$ for some $p >
           0$. Note that
           Equation~\eqref{eq:multibody:multibody_farfield_expansion}
           includes the $n = 2$ and $n = 3$ cases.
           {\small
\begin{align}
& \widetilde{\Phi}(x ; P_2 \times \cdots \times P_n; F( c_{P_2} \times \cdots \times c_{P_n}, p ) ) \notag \\ 
= & 
\prod\limits_{2 \leq s < t \leq n}
\prod\limits_{2 \leq
    k \leq n} 
\sum\limits_{  | \boldsymbol \alpha | \leq p }
\sum\limits_{  \boldsymbol \alpha_{1, k} \geq 0 }
\sum\limits_{ \boldsymbol\alpha_{s, t} \geq 0 }
\sum\limits_{ \boldsymbol\beta_{s, t} \leq \boldsymbol\alpha_{s, t} }
b_k \ M_{ \boldsymbol\xi_k }(P_k, c_{P_k}) \ (-1)^{ \boldsymbol\beta_{s,t}} \notag 
\\
& D^{\boldsymbol\alpha_{1,k}}
  \phi_{1,k} ( x - c_{P_k} )
 D^{ \boldsymbol\alpha_{s, t} } \phi_{s,t} ( c_{P_s} - c_{P_t} )
\label{eq:multibody:truncated_multibody_farfield_expansion}
\end{align}
}

{\noindent \bf Computational Cost of Evaluating the Far-field
  Expansion. }The summations over $\boldsymbol\alpha_{r,s}$ for $1
\leq r < s \leq n$ collectively contribute $\mathcal{O}(p^3)$ terms,
and each inner summation over $\boldsymbol \beta_{s, t}$ contributing
at most $\mathcal{O}(p^3)$ terms. Thus, evaluating the $p$-th order
far-field expansion for a general multibody potential of the form
Equation~\eqref{eq:multibody_potential_form} on a single point takes
$\mathcal{O} \left ( p^{3 \left ( \binom{n - 1}{2} + 1 \right ) }
\right )$ time. In practice, we are forced to use $p = 0$ for $n > 2$
unless most $\phi_{p, q}(x_{i_p}, x_{i_q})$'s in
Equation~\eqref{eq:multibody_potential_form} are constant functions.

\begin{figure}[t]
\centering
\scalebox{0.75}{ \includegraphics{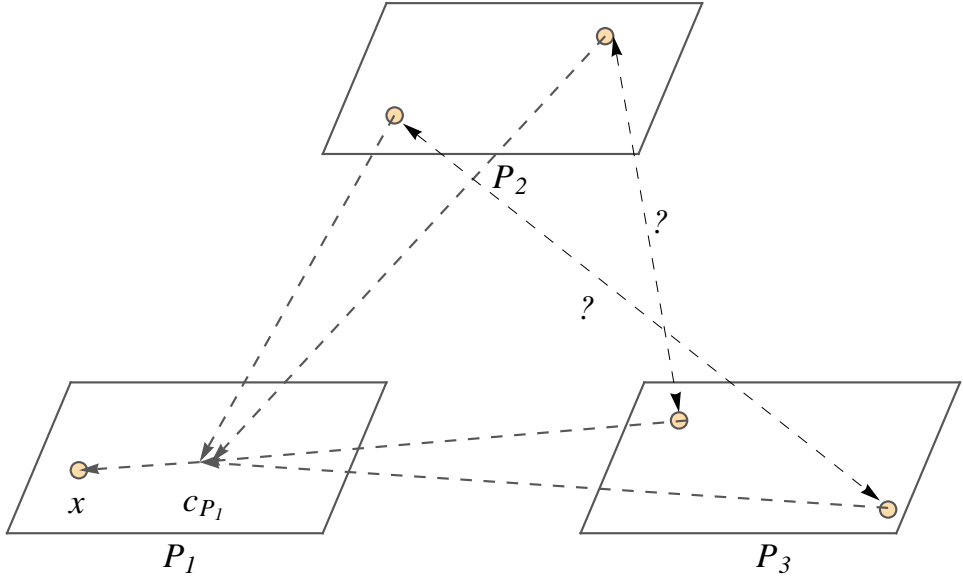}}
\caption{A local expansion created inside the node $P_1$ at $x$ by
  directly accumulating each point in $P_2$ and $P_3$ (see
  Equation~\eqref{eq:multibody:three_body_local_expansion}). We are
  not aware of a technique to express an interaction between a
  particle in $P_2$ and a particle in $P_3$ (marked by the $?$ symbol)
  for $p > 0$.
\label{fig:multibody:three_body_local_expansion}
}
\end{figure}

\subsection{Local Expansion for Three-body Potentials ($n = 3$)}
\label{sec:local_expansion_three_body_potential}
Unlike the far-field expansion case, we are presented a fundamental
difficulty. In order to derive a local expansion, we need to express
the influence of each non-evaluation point $x_{i_j}$ on the evaluation
point $x$ at a center near $x$. However, breaking up the interaction
among the non-evaluation points (i.e. $x_{i_j}$'s in the arguments of
$\phi(x, x_{i_1}, \cdots, x_{i_{n - 1}})$) without loss of information
is hard. To see this: take a three-body potential expressible in
products of pairwise functions (see
Figure~\ref{fig:multibody:three_body_local_expansion}). Expanding near
$c_{P_1}$ inside the node $P_1$ yields an expansion valid for $x \in
P_1$: {\small
\begin{align*}
& \Phi(x ; P_2 \times P_3 ) \\ = &
 \sum\limits_{ x_{i_2} \in P_2 }
 \sum\limits_{ x_{i_3} \in P_3 }
\sum\limits_{ \boldsymbol\alpha_{1,2} \geq 0 } \frac{   
D^{\boldsymbol\alpha_{1,2}} \phi_{1,2} (  c_{P_1} - x_{i_2} )
}{\boldsymbol\alpha_{1,2}!} ( x - c_{P_1})^{\boldsymbol\alpha_{1,2}}
 \sum\limits_{ \boldsymbol\alpha_{1,3} \geq 0 } 
\frac{ D^{\boldsymbol\alpha_{1,3}} \phi_{1,3} (  c_{P_1} - x_{i_3} ) }{\boldsymbol\alpha_{1,3}!}
( x - c_{P_1})^{\boldsymbol\alpha_{1,3}} \\
& \sum\limits_{ \boldsymbol\alpha_{2, 3} \geq 0 }
\frac{ D^{\boldsymbol\alpha_{2,3}} \phi_{2,3} ( c_{P_1} - x_{i_3} ) }{\boldsymbol\alpha_{2,3}!} ( x_{i_2} - c_{P_1})^{\boldsymbol\alpha_{2,3}}
\end{align*}
}Again, let $\boldsymbol \alpha = \boldsymbol \alpha_{1,2} +
\boldsymbol \alpha_{1,3} + \boldsymbol \alpha_{2,3}$. Switching the
orders of summations results: {\small
\begin{align}
  \Phi(x ; P_2 \times P_3 )  = &
\sum\limits_{ \boldsymbol\alpha \geq 0} \
\sum\limits_{ \boldsymbol\alpha_{1,2} \leq \boldsymbol \alpha } \
\sum\limits_{ \boldsymbol\alpha_{1,3} \leq \boldsymbol \alpha  - \boldsymbol \alpha_{1,2} }
\left [  \sum\limits_{ x_{i_2} \in P_2 }
\frac{   D^{\boldsymbol\alpha_{1,2}} \phi_{1,2} (  c_{P_1} - x_{i_2} )
}{\boldsymbol\alpha_{1,2}!}
( x_{i_2} - c_{P_1})^{\boldsymbol\alpha_{2,3}}  \right ] \notag \\
&  \left [ \sum\limits_{ x_{i_3} \in P_3 } 
 \frac{ D^{\boldsymbol\alpha_{1,3}} \phi_{1,3} (  c_{P_1} - x_{i_3} ) }{\boldsymbol\alpha_{1,3}!}
\frac{ D^{\boldsymbol\alpha_{2,3}} \phi_{2,3} ( c_{P_1} - x_{i_3} ) }{\boldsymbol\alpha_{2,3}!}  \right ] 
 ( x - c_{P_1})^{\boldsymbol\alpha_{1,2} + \boldsymbol\alpha_{1,3}} \notag \\
=& 
\sum\limits_{ \boldsymbol\alpha \geq 0}  \left [
\sum\limits_{ \boldsymbol\alpha_{1,2} \leq \boldsymbol \alpha } \
\sum\limits_{ \boldsymbol\alpha_{1,3} \leq \boldsymbol \alpha  - \boldsymbol \alpha_{1,2} }
\bar{N}_{ \boldsymbol\alpha} ( P_2, c_{P_1} ) \ 
\bar{N}_{ \boldsymbol \alpha } ( P_3, c_{P_1} ) \right ]
 ( x - c_{P_1})^{\boldsymbol\alpha_{1,2} + \boldsymbol\alpha_{1,3}} 
\label{eq:multibody:three_body_local_expansion}
\end{align}
}We need the exponent of $(x - c_{P_1})$ to match $\alpha$ to be able
to define the local moments inside $P_1$. Unless $\alpha_{2, 3} = 0$
(i.e. ignore the interaction between a particle in the second set and
a particle in the third set), this is not possible. Since we encounter
a similar problem in the general case, we will skip its discussion.

\section{Simpler Algorithm for General Multibody Potentials}
\label{sec:three_body_algorithms}
Instead of trying to derive the full-fledged tools for general
multibody potentials, we focus on deriving something simpler. Let us
focus on the $n = 3$ case. For a given set of three pairwise disjoint
nodes: $P_1$, $P_2$, $P_3$ and a monotonically
decreasing\footnote{``Monotonic'' multibody potentials decrease in
  value if one of the Euclidean distance arguments is increased while
  the other two are held constant.} three-body potentials such as
$\phi(x_1, x_2, x_3) = \frac{1}{||x_1 - x_2 ||^{\nu_{1, 2}} || x_1 -
  x_3 ||^{ \nu_{1, 3}} || x_2 - x_3 ||^{ \nu_{2, 3} } }$,
\begin{align*}
\forall x_i \in P_1, & 
\widetilde{\Phi}(x_i ; P_2 \times P_3 ) = | P_2 | |
P_3 | \phi( c_{P_1}, c_{P_2}, c_{P_3} )\\
\forall x_j \in P_2, &
\widetilde{\Phi}(x_j ; P_1 \times P_3 ) = | P_1 | | P_3 | \phi (  c_{P_1}, c_{P_2}, c_{P_3} )\\
\forall x_k \in P_3, & \widetilde{\Phi}(x_k ; P_1 \times P_2 ) = | P_1 | | P_2 | \phi ( c_{P_1}, c_{P_2}, c_{P_3}  )
\end{align*}
which can be obtained by setting $p = 0$ in
Equation~\eqref{eq:multibody:truncated_three_body_farfield_expansion}. 
This
means that we can get a cheaper approximation using the number of
points owned by each node. Using the pairwise minimum and maximum node
distances yields: {\small
\begin{align*}
\phi(d^u(P_1, P_2), d^u(P_1, P_3), d^u(P_2, P_3)) \leq  \phi ( c_{P_1}, c_{P_2}, c_{P_3} )  \leq  \phi( d^l(P_1, P_2 ), d^l(P_1, P_3 ), d^l (P_2, P_3 ))
\end{align*}
}It is straightforward to generalize this for the $n \geq 2$ case.

{\noindent \bf Non-monotonic Potentials:} For non-monotonic
potentials such as the Lennard-Jones potential $\phi(x_1, x_2) =
\frac{a}{r^{12}} - \frac{b}{r^6} $, we can compute the critical points
of $\phi$ and determine the intervals of monotonicity of $\phi$ and
consider how $\phi$ behaves in the distance bound range between
$d^l(P_1, P_2)$ and $d^u(P_1, P_2)$. We take a simpler approach that
results in an algorithm that is easier to code; we break up the
potential into two parts such that $\phi(x_1, x_2, \cdots, x_n) =
\phi^+(x_1, x_2, \cdots, x_n) - \phi^-(x_1, x_2, \cdots, x_n)$, and
get a lower and upper bound (though a looser bound) on the
contributions from the positive potential $\phi^+$ and negative
potential $\phi^-$.

\subsection{Specifying the Approximation Rules}
The overall algorithm which also subsumes the pairwise potential case
($n = 2$) was shown in
Algorithm~\ref{alg:multibody:mtpotentialmain}. We can now specify the
$\mbox{\cansummarize}$ function for the general multibody case. For
guaranteeing $\tau$ absolute error bound criterion
(Definition~\ref{defn:bound_absolute_error}), the
$\mbox{\cansummarize}$ function returns true if: $$ \left |
\phi(d^u(P_1, P_2), \cdots, d^u(P_{n -1 }, P_n)) - \phi( d^l(P_1, P_2
), \cdots, d^l (P_{n - 1}, P_n )) \right | \leq \frac{\tau
}{T^{\mathit{root}}} $$ where $T^{root} = \binom{N - 1}{n - 1}$
(i.e. the total number of tuples in each slice in
Figure~\ref{fig:multibody:tensor_slice_sum}).  Let us also define
$T_i$ to be the number of tuples containing a fixed particle in $P_i$ (see Figure~\ref{fig:num_tuples}).
For example, for $n = 3$, the corresponding $\mbox{\summarize}$
function would accumulate for each node:\\ for $P_1$: $| P_2 | | P_3 |
\phi( c_{P_1}, c_{P_2}, c_{P_3} )$, for $P_2$: $| P_1 | | P_3 | \phi (
c_{P_1}, c_{P_2}, c_{P_3} )$, and for $P_3$: $| P_1 | | P_2 | \phi (
c_{P_1}, c_{P_2}, c_{P_3} )$.

{\noindent \bf Hybrid Absolute/Relative Error Guarantee. }The
algorithm for guaranteeing the hybrid absolute/relative error bound
(Definition~\ref{defn:bound_relative_error}) deterministically ($\alpha
= 0$) is not so much different from that for guaranteeing the absolute
error bound. In each node $P$, we maintain the lower bound on the
accumulated potentials for the particles in $P$ (denoted as
$\Phi^l(P)$, a {\it summary statistic} stored in $P$).  The function
$\mbox{\cansummarize}$ returns true if,
\begin{align}
& \left | \phi(d^u(P_1, P_2), \cdots, d^u(P_{n - 1}, P_n)) - \phi(
d^l(P_1, P_2 ), \cdots, d^l (P_{n - 1}, P_n )) \right | \notag \\
\leq &
\frac{\epsilon \min\limits_{1 \leq i \leq n}  ( \Phi^l(P_i) + \delta^l(P_i ; P_1 \times \cdots \times P_{i - 1} \times P_{i + 1} \times \cdots \times P_n) ) + \tau}{T^{\mathit{root}}}
\label{eq:multibody:deterministic_prune}
\end{align}
where each $\delta^l(P_i ; P_1 \times \cdots \times P_{i - 1} \times
P_{i + 1} \times \cdots \times P_n) = \prod\limits_{1 \leq j \leq n, j \neq i} | P_j | \phi(d^u(P_1, P_2),
\cdots, d^u(P_{n -1 }, P_n))$ (which is computed just using the
contribution of the other nodes on the $i$-th node) is added to the
currently running lower bound on each node $\Phi^l(P_i)$ to reflect
the most recently available information on the lower bound.
$\Phi^l(P_i)$ can be incremented and tightened as the computation
progresses, either in the base case or when the recursive sub-calls in
Algorithm~\ref{alg:multibody:mtpotentialmain} are completed (Line 11).

\begin{figure}[t]
\scalebox{0.6}{
\includegraphics[trim=3em 0em 0em 0em,clip=true]{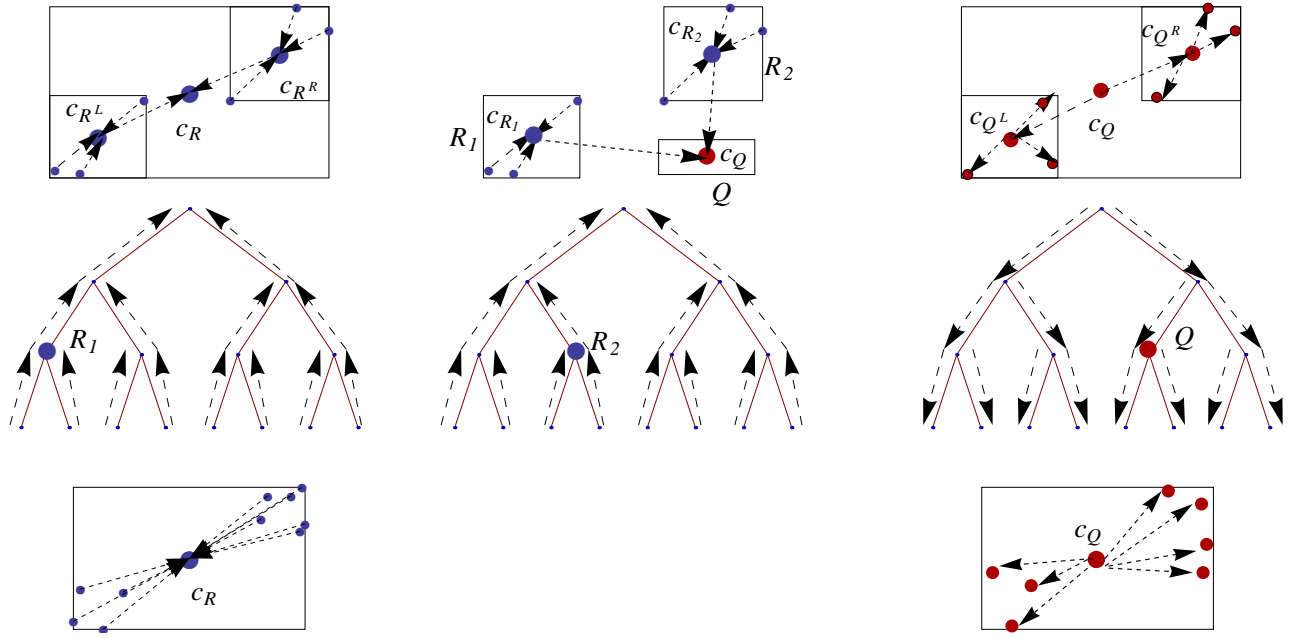}}
\caption{Three-body multipole methods for $p = 0$ in a nutshell.}
\end{figure}

{\noindent \bf Monte Carlo-based Approximations. } The error bounds
provided by the bounding boxes (see Figure~\ref{fig:distance_bounds})
assume that all pairs of points selected between the two nodes are
collapsed to two positions that achieve the minimum distance (and vice
versa for the maximum distance); therefore, these bounds are very
pessimistic and loose. Here we introduce a method for approximating
the potential sums with a probabilistic bound satisfying
Definition~\ref{defn:bound_probabilistic_relative_error}. We can trade
determinism for further gain in efficiency. We have an additional
parameter $\alpha$ that controls the probability level at which the
deviation between each approximation and its corresponding exact
values holds. This was introduced first in~\cite{holmes2010fast,
  holmes2008ultrafast} for probabilistic approximations of aggregate
sums and later extended in~\cite{lee2009fhd} to handle per-particle
quantities.  The theorem that we rely on for probabilistic
approximation is the following:
\begin{thm}
{\bf Central limit theorem: } Let $f_1$, $f_2$, $\cdots$, $f_m$ be
independent, identically distributed samples from the probability
distribution $F$ with variance $\sigma^2$, and $\widetilde{\mu} =
\frac{1}{m} \sum\limits_{s = 1}^m f_s$ be the sample mean of the
samples. As $m \rightarrow \infty$, $\widetilde{\mu} \leadsto N(\mu,
\sigma^2 / m)$.
\end{thm}
A widely accepted statistical rule of thumb asserts that 30 or more
samples are usually enough to put a sample mean into the asymptotic
regime. Berry-Esseen theorem characterizes the rate at which this
convergence to normality takes place more precisely:
{\small
\begin{algorithm}[t]

\caption{$\mbox{\cansummarize} (\{P_i\}_{i=1}^n)$: the Monte Carlo-based approximation.}

\begin{algorithmic}

\IF{$\zeta \cdot m_{\mathit{limit}} \leq \min \{ T_1 , T_2, T_3
  \}$}

\FOR{each $P_i \in \{ P_i \}_{i=1}^n$}

\IF{$i == 1$ or $P_i \not = P_{i - 1}$}

\FOR{$x_{i} \in P_i$}

\IF{$\mbox{\cansummarizemcpoint}(x_i, i, \{P_i\}_{i=1}^n ) ==  \mathbf{false}$}

\RETURN{$\mathbf{false}$}

\ENDIF

\ENDFOR

\ENDIF

\ENDFOR

\RETURN{$\mathbf{true}$}

\ELSE

\RETURN{$\mathbf{false}$}

\ENDIF

\end{algorithmic}

\label{alg:cansummarizemc}
\end{algorithm}
}
\begin{thm}
{\bf Berry-Esseen theorem: }Let $\widetilde{\mu}$ be the sample mean
of $m$ samples drawn from the distribution $F$, and let $\mu$,
$\sigma^2$, and $\rho$ be the mean, variance, and third central moment
of $F$. Let $F_m(x)$ be the cumulative distribution function of
$\widetilde{\mu}$, and $\Psi(x; \mu, \sigma^2)$ be the cdf of the
Gaussian with mean $\mu$ and variance $\sigma^2$. Then there exists a
positive constant $C$ such that for all values of $\widetilde{\mu}$
and $m$:
$$\left | F_m(\widetilde{\mu}) - \Psi(\widetilde{\mu}; \mu, \sigma^2)
\right | \leq \frac{C \rho}{\sigma^3 \sqrt{m}} $$
\end{thm}
\noindent which roughly says that the discrepancy between the normal
distribution and the sample mean distribution goes down as
$\frac{1}{\sqrt{m}}$. For three-body potentials, suppose we are given
the set of three nodes, $P_1$, $P_2$, and $P_3$. Let us consider
$x \in P_1$ (similar approximations can be made for each point in
$P_2$ and $P_3$), and the contribution of $P_2$ and $P_3$ to its
potential sum: { \small
\begin{align*}
\Phi(x ; P_2 \times P_3) = 
\sum\limits_{ x_{i_2} \in X \backslash \{ x \} } \ \sum\limits_{
  \substack{x_{i_3} \in X \backslash \{ x \}\\ i_2 < i_3} }
 \phi \left ( x, x_{i_2}, x_{i_3} \right )
\end{align*}
} We can sample $m$ potential values $\phi(x_{i_1}, x_{i_2}, x_{i_3})$
from the empirical distribution $F$ formed by the 3-tuples formed
among $S_1$, $S_2$, and $S_3$ that contain $x$ in the list. From the
$m$ samples, we get the empirical distribution $F_m^{x}$, from which
we form an approximate $\widetilde{\Phi}(x ; P_2 \times P_3 )$: { \small
$$\widetilde{\Phi}(x ; P_2 \times P_3 ; F_m^{x} ) = T_1
  \widetilde{\mu}_{F_m^{x_i}} = \frac{ T_1 }{ m } \sum\limits_{s =
    1}^m \phi( x_{i_1^s}, x_{i_2^s}, x_{i_3^s} )
$$ }\noindent where $x_{i_1^s} = x$ for all $1 \leq s \leq m$. For
sufficiently large values of $m$, we can assume that the discrepancy
provided by the Berry-Esseen theorem is small and concentrate on the
sample variance of the sample mean distribution. The sample variance
of the sample mean distribution $\widetilde{\sigma}_{\mu_{F_m^{x_i}}}$
is given by:

{\small
\begin{algorithm}[t]

\caption{$\mbox{\cansummarizemcpoint}(x, i, \{S_i\}_{i=1}^n )$: Pruning function for the Monte Carlo based approximation per each point.}

\begin{algorithmic}

\STATE{$F^x \leftarrow \emptyset$}

\REPEAT

\STATE{Get a random $n$-tuple $(x_{1}, \cdots, x_{i-1}, x, x_{i+1},
  \cdots, x_{n})$ where $x_j \in S_j$}

\STATE{$F^x \leftarrow F^{x} \cup \{ \phi(x_{j}, \cdots, x_{j-1}, x,
  x_{j+1}, \cdots, x_{n} ) \}$}

\UNTIL{($ z_{\alpha / 2} \widetilde{\sigma}_{\mu_{F^x}}
  \leq \frac{\tau}{ T^{\mathit{root}} }$ and $|F^x| \geq 30$) or
  $|F^x| \geq m_{\mathit{limit}}$}

\RETURN{$z_{\alpha / 2} \widetilde{\sigma}_{\mu_{F^x}}
  \leq \frac{\tau}{ T^{\mathit{root}} }$}

\end{algorithmic}

\label{alg:cansummarizemcpoint}
\end{algorithm}
}
{\small
$$\widetilde{\sigma}_{\mu_{F_m^{x_i}}} =
\frac{\widetilde{\sigma}_{F_m^{x_i}}}{\sqrt{m}} = \frac{1}{ \sqrt{m} }
\sqrt{ \frac{1}{m - 1} \sum\limits_{s=1}^m ( \phi( x_{i_1^s},
  x_{i_2^s}, x_{i_3^s} ) - \widetilde{\mu}_{F_m^{x_i}} )^2 }$$} where
$\widetilde{\sigma}$ is the sample variance. Given $m$ i.i.d. samples,
with probability of at least $(1 - \alpha)$,
{\small
$$\left | \widetilde{\Phi}(x ; P_2 \times P_3 ; F_m^{x} ) - \Phi(x ; P_2 \times P_3 ) \right | \leq T_1 z_{ \alpha / 2}
\widetilde{\sigma}_{\mu_{F_m^{x_i}}}
$$}where $z_{ \alpha / 2}$ is the number of standard deviations on
either side of $\widetilde{\mu}_{F_m^{x_i}}$ to give at least $(1 -
\alpha)$ coverage under the normal distribution.

{\noindent \bf Modifications to the algorithm. }A Monte Carlo sampling
based routine is shown in Algorithm~\ref{alg:cansummarizemc}. The
function $\mbox{\cansummarize}$ determines whether performing Monte
Carlo approximations (which involves iterating over each unique point
$x \in \bigcup\limits_{i = 1}^n P_i$) with at least
$m_{\mathit{limit}}$ samples is computationally cheaper than the
brute-force computation. $\zeta$ is a global variable that dictates
the desired amount of speedup needed for applying Monte Carlo
approximations, rather than recursing to smaller subsets of the three
nodes. If a desired speedup could be achieved, it loops for each
unique point in $x \in \bigcup\limits_{i = 1}^n P_i$ and computes the
sample mean of the potential values of the tuples that contain $x$,
and the corresponding variance of the sample mean until (1) the
desired error is achieved; or (2) exceeds the number of trial samples
$m_{\mathit{limit}}$. Algorithm~\ref{alg:cansummarizemc} is the form
used for bounding the absolute error of each potential sum error by
$\tau$ with at least probability of $(1 - \alpha)$. For bounding the
hybrid absolute/relative error with at least probability of $(1 -
\alpha)$ (Definition~\ref{defn:bound_probabilistic_relative_error}),
we replace the termination condition in the loop: $z_{\alpha / 2}
\widetilde{\sigma}_{\mu_{F^x}} \leq \frac{\tau}{T^{\mathit{root}}} $
with: {\small
\begin{align}
T_i \cdot z_{\alpha / 2} \widetilde{\sigma}_{\mu_{F^{x_i}}} \leq
\frac{\epsilon ( \Phi^l(P_i) + T_i ( \widetilde{\mu}_{F^{x_i}}
  -z_{\alpha / 2} \widetilde{\sigma}_{\mu_{F^{x_i}}} ) ) + \tau T_i }{
  T^{\mathit{root}} }
\label{eq:prob_rel_error_termination}
\end{align}
}

{\small
\begin{algorithm}[t]

\caption{$\mbox{\summarizemc}(\{S_i\}_{i=1}^n, \{ T_i
\}_{i = 1}^n, \beta)$: Monte Carlo based approximation.}

\begin{algorithmic}

\FOR{each $S_i \in \{ S_i \}_{i=1}^n$}

\IF{$i == 1$ or $S_i \not = S_{i - 1}$}

\FOR{$x_i \in S_i$}

\STATE{$\widetilde{\Phi}(x_i) \leftarrow \widetilde{\Phi}(x_i) + T_i
  \cdot \widetilde{\mu}_{F^{x_i}}$, $\Phi^l(x_i) \leftarrow
  \Phi^l(x_i) + T_i \cdot \left( \widetilde{\mu}_{F^{x_i}} - 
z_{\beta / 2} \widetilde{\sigma}_{\mu_{F^{x_i}}}
\right )$}

\ENDFOR

\ENDIF

\ENDFOR

\end{algorithmic}

\label{alg:summarizemc}
\end{algorithm}
}

\section{Correctness of the Algorithm}
\label{sec:correctness_of_the_algorithm}
The correctness of our algorithm for the deterministic hybrid
absolute/relative error criterion is given by:
\begin{thm}
Algorithm~\ref{alg:multibody:mtpotentialmain} with the function
$\mbox{\cansummarize}$ with the relative error bound guarantee
(Equation~\ref{eq:multibody:deterministic_prune}) produces
approximation $\widetilde{\Phi}(x_{i_1})$ for $x_{i_1} \in X$ such
that
\begin{align}
| \widetilde{\Phi}(x_{i_1}) - \Phi(x_{i_1}) | \leq \epsilon
\Phi(x_{i_1}) + \tau
\label{eq:relerror}
\end{align}
\begin{proof}
(By mathematical induction) For simplicity, let us focus on $n =
  3$. We induct on the number of points $| P_1 \cup P_2 \cup P_3 |$
  encountered during the recursion of the algorithm.

{\noindent \bf Base case: } There are two parts to this part of the proof.
\begin{itemize}
\item{Line 1 of the function $\mbox{\mtpotentialcanonical}$ in
Algorithm~\ref{alg:multibody:mtpotentialmain}: any set of nodes $P_1$,
$P_2$, $P_3$ for which the function $\mbox{\cansummarize}$ returns
true satisfies the error bounds for $x_{i_u} \in S_u$ for $u = 1, 2,
3$:
{\small
\begin{align}
\forall x_{i_1} \in P_1, & \left | \widetilde{\Phi}(x_{i_1} ; P_2 \times P_3 ) - \Phi(x_{i_1} ; P_2 \times P_3 ) \right | \notag \\ 
\leq & \frac{T_{x_{i_u} \times P_2 \times P_3}}{T^{ \mathit{root} }} \left ( \epsilon \Phi^l(P_1)  + \tau \right )
\leq \frac{T_{x_{i_u} \times P_2 \times P_3}}{T^{ \mathit{root} }} \left ( \epsilon \Phi(x_{i_1})  + \tau \right ) \notag \\
\forall x_{i_2} \in P_2, & \left | \widetilde{\Phi}(x_{i_2} ; P_1 \times P_3 ) - \Phi(x_{i_2} ; P_1 \times P_3 ) \right | \notag \\
\leq & \frac{T_{x_{i_u} \times P_2 \times P_3} }{T^{ \mathit{root} }} \left ( \epsilon \Phi^l(P_2)  + \tau \right )
\leq \frac{T_{x_{i_u} \times P_2 \times P_3}}{T^{ \mathit{root} }} \left ( \epsilon  \Phi(x_{i_2}) + \tau \right ) \notag \\
\forall x_{i_3} \in P_3, & \left | \widetilde{\Phi}(x_{i_3} ; P_1 \times P_2 ) - \Phi(x_{i_3} ; P_1 \times P_2 ) \right | \notag \\
\leq & \frac{T_{x_{i_u} \times P_2 \times P_3}}{T^{ \mathit{root} }} \left ( \epsilon \Phi^l(P_3)  + \tau \right )
\leq \frac{ T_{x_{i_u} \times P_2 \times P_3}}{T^{ \mathit{root} }} \left ( \epsilon \Phi(x_{i_3}) + \tau \right )
\label{eq:implied_by_pruning_rule}
\end{align}
}where $T_{x_{i_u} \times P_2 \times P_3}$ denotes the number of tuples chosen by fixing $x_{i_u}$ and selecting the other two from $P_2$ and $P_3$ and so on.
}
\item{The function call $\mbox{\mtpotentialbase}$ in 
Algorithm~\ref{alg:multibody:mtpotentialmain}: each $x_{i_1} \in P_1$ and
$x_{i_2} \in P_2$ and $x_{i_3} \in P_3$ exchange contributions exactly
and incur no approximation error.}
\end{itemize}

{\noindent \bf Inductive step: } Suppose we are given the set of three
nodes $P_1$, $P_2$, and $P_3$ (at least one of which is an internal
node) in the function $\mbox{\mtpotentialcanonical}$. Suppose the
three tuples $P_1$, $P_2$, $P_3$ could not be pruned, and that we need
to recurse on each child of $P_1$, $P_2$, and $P_3$.

By assumption, $\mbox{\cansummarize}$ returns false if any one of the
nodes $P_1$, $P_2$, $P_3$ includes one of the other nodes (see
Section~\ref{sec:generalized_nbody_framework}). For $n = 3$, we can
assume that the possible node tuple cases that could be considered for
pruning are shown in Figure~\ref{fig:num_tuples}. Let $\{ \{ P_{s}^k
\}_{s = 1}^3 \}_{k = 1}^t$ be the set of set of three nodes considered
during the recursive sub-computations using the child nodes of each
$P_1$, $P_2$, and $P_3$; note that the maximum value of $t$ is 8 for
three-body interactions. Note that for each $k$, $P_{s}^k$ is either
(1) the node $P_s$ itself (2) the left child node of $P_s$ (3) the
right child node of $P_s$. Therefore, for each $k = 1, 2, \cdots, t$,
$| P_{1}^k \cup P_{2}^k \cup P_{3}^k | \leq | P_1 \cup P_2 \cup P_3
|$. The equality holds when all of $P_1$, $P_2$, and $P_3$ are leaf
  nodes for which the error criterion is satisfied by the base case
  function (no error incurred).

If any one of $P_1$, $P_2$, and $P_3$ is an internal node, then we are
guaranteed that $| P_{1}^k \cup P_{2}^k \cup P_{3}^k | < | P_1 \cup
P_2 \cup P_3 |$ for all $k = 1, \cdots, t$. We invoke the inductive
hypothesis to conclude that for each $k$ and for each $x_{i_u} \in
P_{u}^k$ for $u = 1, 2, 3$: {\small
\begin{align*}
\forall x_{i_1} \in P_1^k, & \left | \widetilde{\Phi}(x_{i_1} ; P_2^k \times P_3^k ) - \Phi(x_{i_1} ; P_2^k \times P_3^k ) \right |  \notag \\
\leq & \frac{T_{x_{i_1} \times P_2^k \times P_3^k } }{T^{ \mathit{root} }} \left ( \epsilon \Phi^l(P_1^k)  + \tau \right )
\leq \frac{T_{x_{i_1} \times P_2^k \times P_3^k } }{T^{ \mathit{root} }} \left ( \epsilon \Phi(x_{i_1})  + \tau \right ) \notag \\
\forall x_{i_2} \in P_2^k, & \left | \widetilde{\Phi}(x_{i_2} ; P_1^k \times P_3^k ) - \Phi(x_{i_2} ; P_1^k \times P_3^k ) \right | \notag \\
\leq & \frac{T_{x_{i_2} \times P_1^k \times P_3^k }  }{T^{ \mathit{root} }} \left ( \epsilon \Phi^l(P_2^k)  + \tau \right )
\leq \frac{T_{x_{i_2} \times P_1^k \times P_3^k } }{T^{ \mathit{root} }} \left ( \epsilon  \Phi(x_{i_2}) + \tau \right ) \notag \\
\forall x_{i_3} \in P_3^k, & \left | \widetilde{\Phi}(x_{i_3} ; P_1^k \times P_2^k ) - \Phi(x_{i_3} ; P_1^k \times P_2^k ) \right | \notag \\
\leq &\frac{T_{x_{i_3} \times P_1^k \times P_2^k } }{T^{ \mathit{root} }} \left ( \epsilon \Phi^l(P_3^k)  + \tau \right )
\leq \frac{ T_{x_{i_3} \times P_1^k \times P_2^k } }{T^{ \mathit{root} }} \left ( \epsilon \Phi(x_{i_3}) + \tau \right )
\end{align*}
}where $T_{s}^k$ is the number of 3-tuples formed among $P_{1}^k$,
$P_{2}^k$, $P_{3}^k$ that contain a fixed point in $P_{s}^k$.  By the
triangle inequality, Equation~\ref{eq:relerror} holds by extending to
$P_1 = P_2 = P_3 = X$ since the number of encountered tuples for each
particle add up to $T^{\mathit{root}}$.
\end{proof}
\label{thm:rel_error}
\end{thm}
We are now ready to prove the correctness of our algorithm for
bounding the relative error probabilistically.
\begin{thm}
Algorithm~\ref{alg:multibody:mtpotentialmain} with the function
$\mbox{\cansummarize}$ with the modification described in
Equation~\ref{eq:prob_rel_error_termination} produces approximations
$\widetilde{\Phi}(x_{i})$ for $x_i \in X$ such that
\begin{align}
| \widetilde{\Phi}(x_{i}) - \Phi(x_{i}) | \leq \epsilon \Phi(x_{i}) + \tau
\label{eq:probrelerror}
\end{align}
with the probability of at least $1 - \alpha$ for $0 < \alpha < 1$, as
the number of samples in the Monte Carlo approximation tends to
infinity.
\begin{proof}
We extend the proof in Theorem~\ref{thm:rel_error}. For simplicity, we
again focus on the $n = 3$ case.

{\noindent \bf Base case: }Given the set of three nodes with the
desired failure probability $\alpha$, the base case
$\mbox{\mtpotentialbase}$ is easily shown to satisfy
Equation~\ref{eq:implied_by_pruning_rule} with 100 \% probability ( $>
1 - \alpha$). Similarly, each Monte Carlo prune satisfies
Equation~\ref{eq:implied_by_pruning_rule} with probability of $1 -
\alpha$ asymptotically.

{\noindent \bf Inductive case: }For a non-prunable set of three nodes
$\{ P_k \}_{k = 1}^3$ for the required failure probability
$\beta$. Note that $\mbox{\mtpotentialcanonical}$ results in a maximum
of four (i.e. $2^{3 - 1} = 4$) sub-calls for a set of non-prunable
$P_1$, $P_2$, $P_3$ nodes. For example, suppose $P_1$ is an internal
node, and consider its left child, $P_1^L$. The contribution of $P_2$
and $P_3$ on $P_1^L$ can be computed by considering the node
combinations: $(P_1^L, P_2^L, P_3^L)$, $(P_1^L, P_2^L, P_3^R)$,
$(P_1^L, P_2^R, P_3^L)$, $(P_1^L, P_2^R, P_3^R)$, resulting in a
maximum of four combinations if $P_1$, $P_2$, $P_3$ satisfy the
case~\ref{fig:num_tuples_canonical1} in
Figure~\ref{fig:num_tuples}. Each recursive sub-call is equivalent to
a stratum in a stratified sampling, and satisfies the following:
{\small
\begin{align*}
 \left | \widetilde{\Phi}(x_{i_u}; P_2^L \times P_3^L) -
  \Phi (x_{i_u} ; P_2^L \times P_3^L)  \right | & \leq
  \frac{\epsilon T_{ x_{i_u} \times P_2^L \times P_3^L } }{T^{\mathit{root}}} \Phi^l(P_1^L) + \frac{ \tau T_{ x_{i_u} \times P_2^L \times P_3^L }}{T^{\mathit{root}}} \\
 \left | \widetilde{\Phi}(x_{i_u}; P_2^L \times P_3^R) -
  \Phi (x_{i_u} ; P_2^L \times P_3^R)  \right | & \leq
  \frac{\epsilon T_{ x_{i_u} \times P_2^L \times P_3^R } }{T^{\mathit{root}}} \Phi^l(P_1^L) + \frac{ \tau T_{ x_{i_u} \times P_2^R \times P_3^L }}{T^{\mathit{root}}} \\
 \left | \widetilde{\Phi}(x_{i_u}; P_2^R \times P_3^L) -
  \Phi (x_{i_u} ; P_2^R \times P_3^L)  \right | & \leq
  \frac{\epsilon T_{ x_{i_u} \times P_2^R \times P_3^L } }{T^{\mathit{root}}} \Phi^l(P_1^L) + \frac{ \tau T_{ x_{i_u} \times P_2^R \times P_3^R }}{T^{\mathit{root}}} \\
 \left | \widetilde{\Phi}(x_{i_u}; P_2^R \times P_3^R) -
  \Phi (x_{i_u} ; P_2^R \times P_3^R)  \right | & \leq
  \frac{\epsilon T_{ x_{i_u} \times P_2^R \times P_3^R } }{T^{\mathit{root}}} \Phi^l(P_1^L) + \frac{ \tau T_{ x_{i_u} \times P_2^R \times P_3^R }}{T^{\mathit{root}}} 
\end{align*}
}
Collectively, the results from these strata add up to potential
estimates that satisfy the error bound with at least $1 - \alpha$
probability for each $x_{i_u} \in P_1^L$ and the following holds:
{\small
\begin{align*}
 \left | \widetilde{\Phi}(x_{i_u}; P_2 \times P_3) -
  \Phi (x_{i_u} ; P_2 \times P_3)  \right | & \leq
  \frac{\epsilon T_{ x_{i_u} \times P_2 \times P_3 } }{T^{\mathit{root}}} \Phi^l(x_{i_u}) + \frac{ \tau T_{ x_{i_u} \times P_2 \times P_3 }}{T^{\mathit{root}}} 
\end{align*}
}where $T_{ x_{i_u} \times P_2 \times P_3 } = T_{ x_{i_u} \times P_2^L
  \times P_3^L } + T_{ x_{i_u} \times P_2^L \times P_3^R } + T_{
  x_{i_u} \times P_2^R \times P_3^L } + T_{ x_{i_u} \times P_2^R
  \times P_3^R }$. The similar bounds hold for each $ x \in P_1^R$,
and the same reasoning can be extended to the bounds for $P_2$ and
$P_3$.  Because $\Phi^l(P_1) = \min \{ \Phi^l(P_1^L), \Phi^l(P_1^R) \}
$ throughout the execution of the algorithm, we can extend the
argument to the case where $P_1 = P_2 = P_3 = X$.
\end{proof}
\end{thm}

\begin{figure}[t]
\centering
\includegraphics[width=0.8\textwidth]{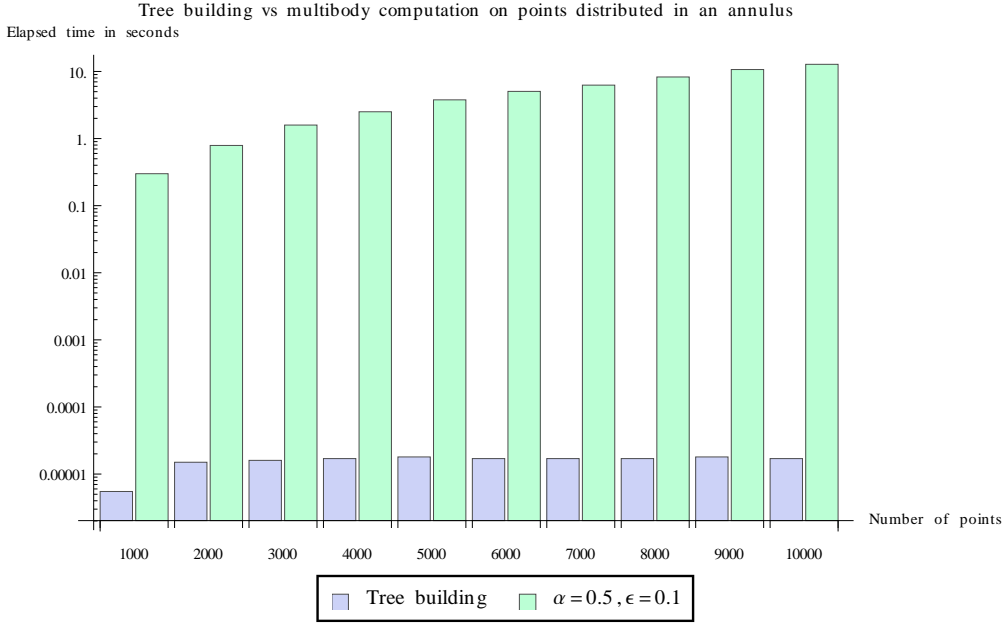}
\caption{Building the $kd$-tree takes negligible amount of time
  compared to the time it takes for the actual multibody
  computation.\label{fig:multibody:tree_building_multibody_computation}}
\end{figure}

\section{Experiment Results}
\label{sec:experimental_results}
All of our algorithms were based on an open-source C++ library called
MLPACK~\cite{gray2009mlpack,curtin2011libmlpack}. The experiments were
performed on a desktop with AMD Phenom II X6 1100T Processors
utilizing only one core with 8 GB of RAM.

\subsection{Tree Building}
The cost of tree-building is negligible compared to the actual
multibody computation. Compared to complex, irregular memory access
patterns encountered in the multibody computation (as do most
recursive algorithms in general), the tree-building phase requires
mostly sequential scanning of contiguous blocks of memory and thus
requires shorter amount of time. See
Figure~\ref{fig:multibody:tree_building_multibody_computation}, where
the tree building is compared to the multibody computation with the
relative error criterion $\epsilon = 0.1$ and the 50 \% probability
guarantee ($\alpha = 0.5$). The annulus distribution was chosen
deliberately to show that even under the distribution for which the
multibody computation is relatively fast (see
Section~\ref{sec:multibody:multibody_computation}), the tree building
requires a tiny fraction of time compared to the computation time.

\begin{figure}[t]
\centering
\includegraphics[width=0.8\textwidth]{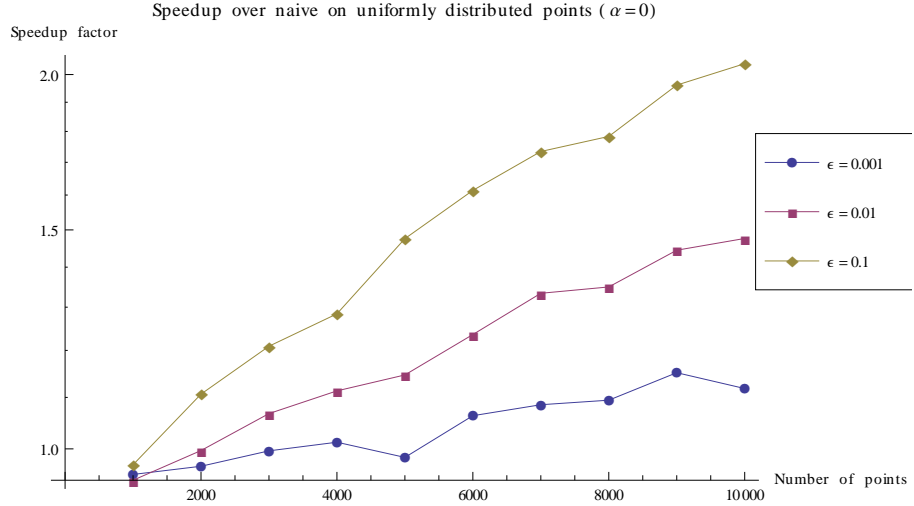}
\caption{Speedup result on uniformly distributed points using the
  deterministic algorithm ($\alpha = 0$). The base timings for the
  naive algorithm on each point set are: $1.91 \times 10^1$ seconds,
  $1.54 \times 10^2$ seconds, $5.17 \times 10^2$ seconds, $1.23 \times
  10^3$ seconds, $2.39 \times 10^3$ seconds, $4.16 \times 10^3$
  seconds, $6.64 \times 10^3$ seconds, $9.76 \times 10^3$ seconds,
  $1.43 \times 10^4$ seconds, and $1.92 \times 10^4$
  seconds.\label{fig:multibody:multibody_uniform_100}}
\end{figure}

\subsection{Multibody Computation}
\label{sec:multibody:multibody_computation}
We demonstrate speedup results of our approximate algorithms
guaranteeing the $(1 - \alpha)$ probabilistic $\epsilon$ relative
error criterion
(Definition~\ref{defn:bound_probabilistic_relative_error}). For this
paper, we focus strictly on the relative error criterion ($\tau = 0$)
and test on three relative error parameter values ( $\epsilon =
0.001$, $\epsilon = 0.01$, and $\epsilon = 0.1$). We test on three
different types of distribution: uniform within the unit hypercube
$[0, 1]^3$ (denoted as the ``uniform" distribution), the annulus
distribution (denoted as the ``annulus" distribution) in three
dimensions, and uniform within the unit three-dimensional sphere
(denoted as the ``ball'' distribution).  These three distributions
were also used in~\cite{bentley1990k}). For the deterministic and
probabilistic algorithms, the order of local expansion is fixed at $p
= 0$ and only $0$-th order multipole expansions are used for the
results.

\begin{figure}[t]
\centering
\includegraphics[width=0.8\textwidth]{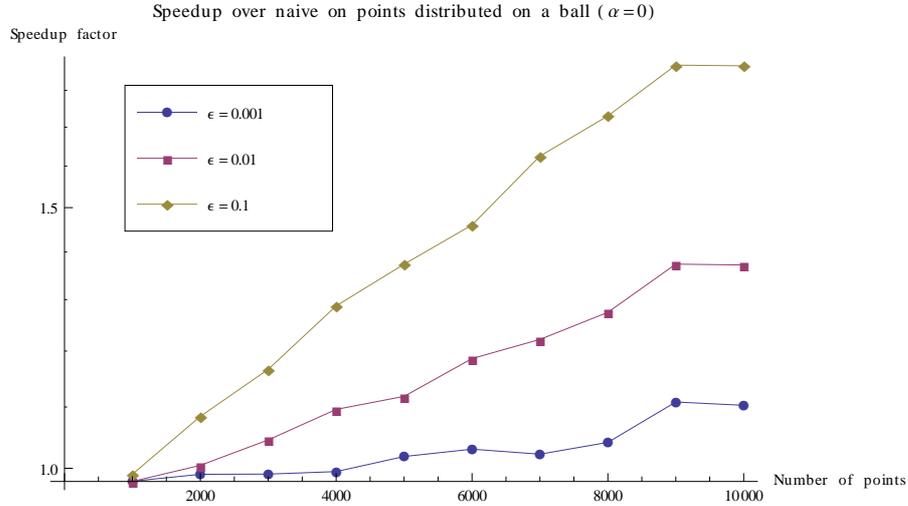}
\caption{Speedup result on points distributed inside a sphere using
  the deterministic algorithm ($\alpha = 0$). The base timings for the
  naive algorithms are listed in
  Figure~\ref{fig:multibody:multibody_uniform_100}.\label{fig:multibody:multibody_ball_100}}
\end{figure}

\begin{figure}
\centering
\includegraphics[width=0.8\textwidth]{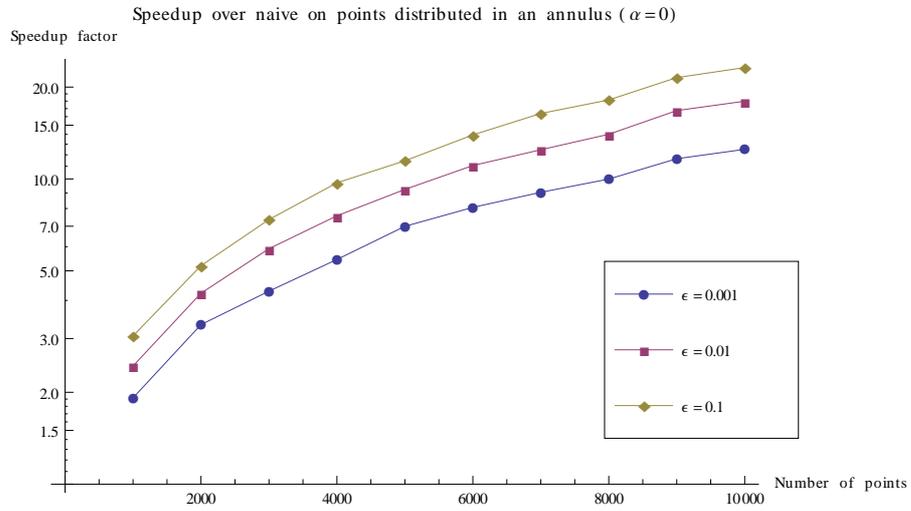}
\caption{Speedup result on points distributed on an annulus using the
  deterministic algorithm ($\alpha = 0$). The base timings for the
  naive algorithms are listed in
  Figure~\ref{fig:multibody:multibody_uniform_100}.\label{fig:multibody:multibody_annulus_100}}
\end{figure}

{\noindent \bf Deterministic
  Approximations. }Figure~\ref{fig:multibody:multibody_uniform_100},
Figure~\ref{fig:multibody:multibody_annulus_100}, and
Figure~\ref{fig:multibody:multibody_ball_100} show speedup results
against the naive algorithm using only the deterministic approximation
(i.e. $\alpha = 0$). On the uniform distribution and the ball
distribution, the speedup is almost non-existent; the speedup factor
is a little bit more than two on the dataset containing $10,000$
points using the lowest parameter setting of $\epsilon = 0.1$. On the
annulus distribution, our deterministic algorithm achieves a little bit
better speedup against the naive algorithm; a factor of more than 20
times speedup on $10,000$ points is encountered on $\epsilon = 0.1$. A
tree-based hierarchical method generally works better for clustered
point sets, and this is reflected in our results.

\begin{figure}
\centering
\includegraphics[width=0.8\textwidth]{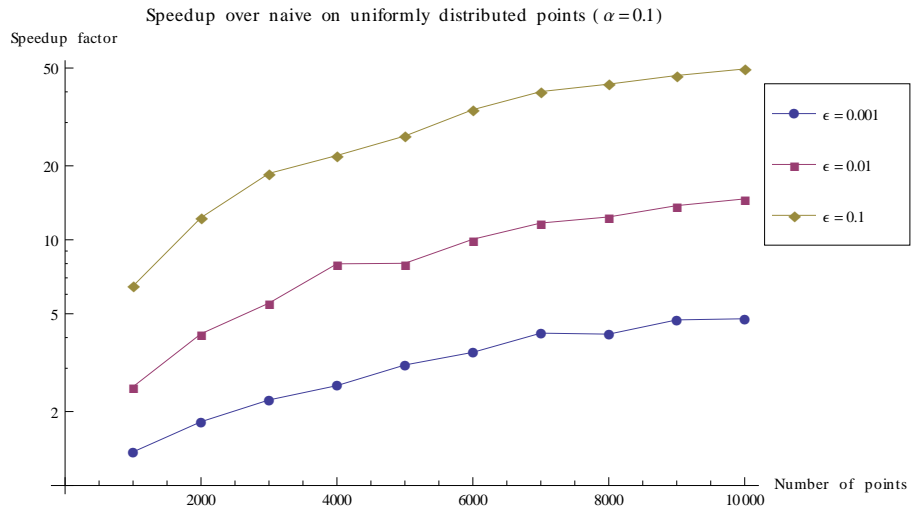}
\caption{Speedup result on uniformly distributed points using the
  Monte Carlo-based algorithm ($\alpha =
  0.1$). The base timings for the
  naive algorithms are listed in
  Figure~\ref{fig:multibody:multibody_uniform_100}.
\label{fig:multibody:multibody_uniform}}
\end{figure}

\begin{figure}
\centering
\includegraphics[width=0.8\textwidth]{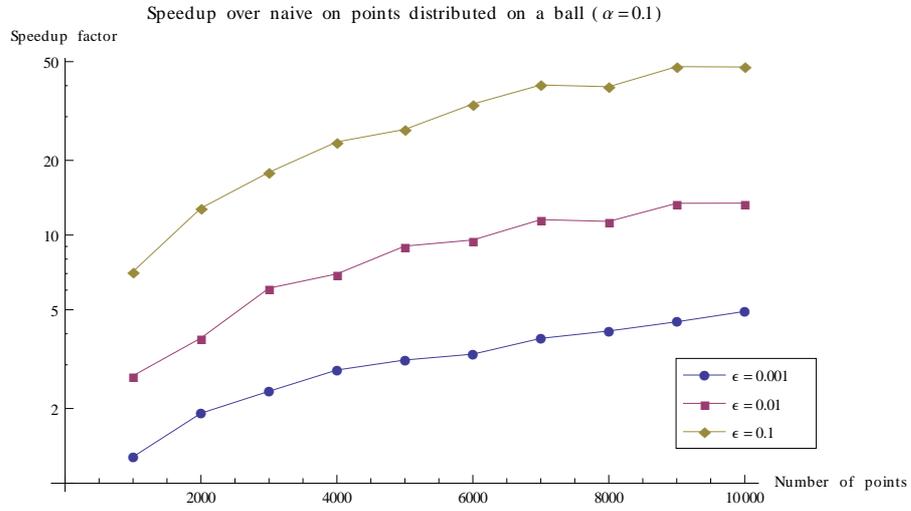}
\caption{Speedup result on points distributed inside a sphere using
  the Monte Carlo-based algorithm ($\alpha = 0.1$). The base timings
  for the naive algorithms are listed in
  Figure~\ref{fig:multibody:multibody_uniform_100}.
\label{fig:multibody:multibody_ball}}
\end{figure}

\begin{figure}
\centering
\includegraphics[width=0.8\textwidth]{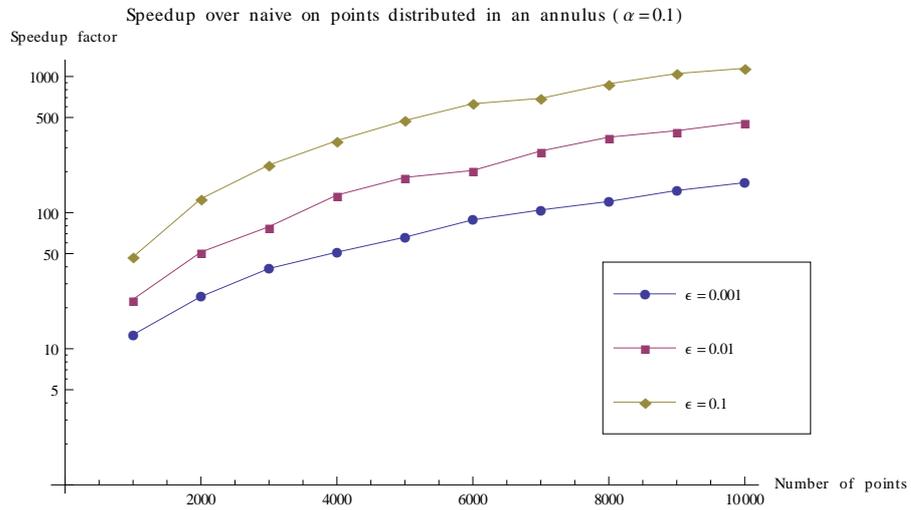}
\caption{Speedup result on points distributed on an annulus using the
  Monte Carlo-based algorithm ($\alpha = 0.1$). The base timings for
  the naive algorithms are listed in
  Figure~\ref{fig:multibody:multibody_uniform_100}.
\label{fig:multibody:multibody_annulus}}
\end{figure}

{\noindent \bf Monte-Carlo Approximations. }In this section, we show
whether adding indeterminism by sampling can reduce the computation
time while guaranteeing a slightly relaxed error criterion (but with a
high probability guarantee for each potential sum).  We first relax
the probability guarantee to be $90 \%$ (i.e. $\alpha = 0.1$). Like
the results shown using the deterministic algorithm, our Monte
Carlo-based algorithm achieves the most speedup on points distributed
in an annulus ($1000$ times speedup on $10,000$ points using $\epsilon
= 0.1$). See Figure~\ref{fig:multibody:multibody_uniform}, Figure~\ref{fig:multibody:multibody_ball}, and Figure~\ref{fig:multibody:multibody_annulus}.

We also list the percentage of the points actually achieving the
$\epsilon$ relative error bound along with the mean and the variance
in Table~\ref{table:multibody:error_distribution_uniform},
Table~\ref{table:multibody:error_distribution_ball}, and
Table~\ref{table:multibody:error_distribution_annulus}. The relative
error level of $0.001$ and the probability guarantee of $90 \%$ was
used. Under all three distributions, the percentage of points whose
potential sum achieved the desired relative error of $0.001$ was well
above $90 \%$. We list the average relative error, the variance, and
the maximum relative error. Note that the maximum relative error can
exceed $100 \%$ if the true potential sum and its approximation have
opposite signs. For a particle with a small potential sum, we have
observed that this is indeed the case due to numerical inaccuracies
accumulated during the summation.
\begin{table}
{\small
\begin{center}
\begin{tabular}{|r|l|l|l|l|}
\hline
\multirow{2}{5em}{Number of points} & \% achieving & \multirow{2}{7em}{Average relative error} & \multirow{2}{7em}{Variance} & \multirow{2}{7em}{Maximum relative error}\\
 & &  &  &  \\
\hline
1000 & $98.3 \%$ & $1.11 \times 10^{-4}$ & $8.89 \times 10^{-7}$ & $2.86 \times 10^{-2}$\\
2000 & $97.9 \%$ & $1.28 \times 10^{-4}$ & $7.71 \times 10^{-7}$ & $2.78 \times 10^{-2}$\\
3000 & $98.6 \%$ & $1.51 \times 10^{-4}$ & $2.64 \times 10^{-6}$ & $6.47 \times 10^{-2}$ \\
4000 & $98.3 \%$ & $1.44 \times 10^{-4}$ & $3.37 \times 10^{-6}$ & $1.01 \times 10^{-1}$ \\
5000 & $98.7 \%$ & $2.65 \times 10^{-4} $ & $ 1.09 \times 10^{-4}$ & $7.36 \times 10^{-1}$ \\
6000 & $98.3 \%$ & $1.29 \times 10^{-4}$ & $1.39 \times 10^{-6}$ & $3.62 \times 10^{-2}$ \\
7000 & $98.4 \%$ & $1.86 \times 10^{-4}$ & $9.29 \times 10^{-6}$ & $1.96 \times 10^{-1}$ \\
8000 & $98.8 \%$ & $9.89 \times 10^{-5}$ & $1.21 \times 10^{-6}$ & $6.50 \times 10^{-2}$\\
9000 & $98.8 \%$ & $9.94 \times 10^{-5}$ & $1.39 \times 10^{-6}$ & $6.69 \times 10^{-2}$ \\
10000 & $98.9 \%$ & $1.02 \times 10^{-4}$ & $1.95 \times 10^{-6}$ & $1.06 \times 10^{-1}$ \\
\hline
\end{tabular}
\end{center}
}
\caption{The distribution of relative error on the uniform distribution
  using $\alpha = 0.1$ and $\epsilon = 0.001$.\label{table:multibody:error_distribution_uniform}}
\end{table}

\begin{table}
{\small
\begin{center}
\begin{tabular}{|r|l|l|l|l|}
\hline
\multirow{2}{5em}{Number of points} & \% achieving & \multirow{2}{7em}{Average relative error} & \multirow{2}{7em}{Variance} & \multirow{2}{7em}{Maximum relative error}\\
 & &  &  &  \\
\hline
1000 & $98.6 \%$ & $8.21 \times 10^{-5}$ & $1.17 \times 10^{-7}$ & $7.22 \times 10^{-3}$\\
2000 & $98.7 \%$ & $1.35 \times 10^{-4}$ & $1.26 \times 10^{-6}$ & $2.78 \times 10^{-2}$\\
3000 & $98.7 \%$ & $1.11 \times 10^{-4}$ & $7.58 \times 10^{-7}$ & $3.23 \times 10^{-2}$ \\
4000 & $97.0 \%$ & $1.36 \times 10^{-3}$ & $1.21 \times 10^{-3}$ & $1.81 \times 10^{0}$ \\
5000 & $98.2 \%$ & $1.19 \times 10^{-4} $ & $1.18 \times 10^{-6}$ & $4.85 \times 10^{-2}$ \\
6000 & $98.9 \%$ & $1.20 \times 10^{-4}$ & $3.70 \times 10^{-6}$ & $1.27 \times 10^{-1}$ \\
7000 & $98.8 \%$ & $1.22 \times 10^{-4}$ & $3.32 \times 10^{-6}$ & $1.11 \times 10^{-1}$ \\
8000 & $98.5 \%$ & $1.31 \times 10^{-4}$ & $3.67 \times 10^{-6}$ & $1.12 \times 10^{-1}$\\
9000 & $97.9 \%$ & $6.24 \times 10^{-4}$ & $3.89 \times 10^{-4}$ & $1.14 \times 10^{0}$ \\
10000 & $97.6 \%$ & $5.09 \times 10^{-4}$ & $2.40 \times 10^{-4}$ & $1.28 \times 10^{0}$ \\
\hline
\end{tabular}
\end{center}
}
\caption{The distribution of relative error on the ball distribution
  using $\alpha = 0.1$ and $\epsilon = 0.001$.\label{table:multibody:error_distribution_ball}}
\end{table}

\begin{table}
{\small
\begin{center}
\begin{tabular}{|r|l|l|l|l|}
\hline
\multirow{2}{5em}{Number of points} & \% achieving & \multirow{2}{7em}{Average relative error} & \multirow{2}{7em}{Variance} & \multirow{2}{7em}{Maximum relative error}\\
 & &  &  &  \\
\hline
1000 & $98.4 \%$ & $9.33 \times 10^{-5}$ & $3.42 \times 10^{-7}$ & $1.38 \times 10^{-2}$\\
2000 & $97.2 \%$ & $9.21 \times 10^{-4}$ & $2.69 \times 10^{-4}$ & $5.15 \times 10^{-1}$\\
3000 & $98.7 \%$ & $8.52 \times 10^{-5}$ & $1.16 \times 10^{-6}$ & $5.09 \times 10^{-2}$ \\
4000 & $91.8 \%$ & $2.53 \times 10^{-2}$ & $6.10 \times 10^{-1}$ & $4.80 \times 10^{1}$ \\
5000 & $96.9 \%$ & $1.28 \times 10^{-3} $ & $1.09 \times 10^{-3}$ & $1.27 \times 10^{0}$ \\
6000 & $92.8 \%$ & $6.28 \times 10^{-3}$ & $1.38 \times 10^{-2}$ & $6.43 \times 10^{0}$ \\
7000 & $95.2 \%$ & $2.13 \times 10^{-3}$ & $1.36 \times 10^{-3}$ & $6.66 \times 10^{-4}$ \\
8000 & $91.2 \%$ & $1.45 \times 10^{-2}$ & $3.77 \times 10^{-1}$ & $5.36 \times 10^{1}$\\
9000 & $94.6 \%$ & $5.17 \times 10^{-3}$ & $6.56 \times 10^{-3}$ & $3.94 \times 10^{0}$ \\
10000 & $91.6 \%$ & $2.72 \times 10^{-2}$ & $8.06 \times 10^{-1}$ & $8.29 \times 10^{1}$ \\
\hline
\end{tabular}
\end{center}
}
\caption{The distribution of relative error on the annulus distribution
  using $\alpha = 0.1$ and $\epsilon = 0.001$.\label{table:multibody:error_distribution_annulus}}
\end{table}

\section{Conclusion}
In this paper, we have introduced the framework for extending the
pairwise series expansion to potentials that involve more than two
points. Through this process, we have formally defined an analogue to
the far-field expansion for approximating the multibody potentials in
a hierarchical fashion as done in traditional FMM algorithms and have
derived algorithms for guaranteeing (1) absolute error bound (2)
relative error bound (3) probabilistic absolute/relative error on each
particle potential sum and proved the correctness of our algorithms
formally. However, we do not present a full-fledged derivation of all
three translation operators and the analogue to the local expansion
due to a technical difficulty. Instead, we propose to use only a
monopole approximation ($p = 0$) in a simpler alternative algorithm.
Our experiment demonstrates that the algorithm using the hybrid
deterministic/probabilistic approximation heuristic achieves speedup
under points lying on an annulus of a sphere (i.e. lower-dimensional
manifold). For our future work, we are working on parallelization as
done in~\cite{li2006mfd,sampath2010pfg}.

\section*{References}
\bibliographystyle{elsarticle-num}
\bibliography{../phd_thesis/thesis_references.bib}

\end{document}

%% file: thesis_macro.tex
\newtheorem{thm}{Theorem}[section]

\newtheorem{defn}[thm]{Definition}

\newcommand{\buildkdtree}{\textsc{BuildKdTree}}

\newcommand{\cansummarize}{\textsc{CanSummarize}}

\newcommand{\cansummarizemcpoint}{\textsc{CanSummarizeMCPoint}}
\newcommand{\summarize}{\textsc{Summarize}}
\newcommand{\summarizemc}{\textsc{SummarizeMC}}

\newcommand{\mtpotentialcanonical}{\textsc{MTPotentialCanonical}}

\newcommand{\mtpotentialbase}{\textsc{MTPotentialBase}}

